\documentclass[a4paper,USenglish,cleveref, autoref, thm-restate]{lipics-v2019}

\hideLIPIcs  
\nolinenumbers

\usepackage{amssymb,amsmath,amsthm}
\usepackage{color}
\usepackage{graphicx}
\usepackage{hyperref}
\usepackage{cite}
\usepackage{tikz}
\usetikzlibrary{matrix}

\newcommand{\m}[1]{\mathcal{#1}}
\newcommand{\local}{\textsf{LOCAL}}
\newcommand{\alg}{\textsc{alg}}
\newcommand{\LMIS}{\m{M}_2}
\newcommand{\LCOL}{\m{C}}
\newcommand{\id}{\mbox{\rm\textsf{id}}}
\DeclareMathOperator{\poly}{poly}

\title{The Topology of Local Computing in Networks}

\authorrunning{P. Fraigniaud and A. Paz} 

\author
{Pierre Fraigniaud}
{Institut de Recherche en Informatique Fondamentale, CNRS and Universit\'e Paris Cit\'e, France}
{pierre.fraigniaud@irif.fr }
{}
{Supported by ANR Project DUCAT (ref.~ANR-20-CE48-0006)}

\author
{Ami Paz}
{Laboratoire Interdisciplinaire des Sciences du Num\'erique, CNRS and Universit\'e Paris-Saclay, France}
{ami.paz@lisn.fr}
{0000-0002-6629-8335}
{}

\Copyright{Pierre Fraigniaud and Ami Paz} 



\acknowledgements{We are grateful to the reviewers of the Journal of Applied and Computational Topology for their valuable comments}

\relatedversion{A preliminary version of this work was presented in ICALP 2020}

\begin{document}

%

\maketitle

\begin{abstract}
For more than three decades, distributed systems have been described and analyzed using topological tools, primarily using two techniques: protocol complexes and directed algebraic topology.
In both cases, the considered computational model generally assumes communication via shared objects (typically a shared memory consisting of a collection of read-write registers) or message-passing enabling direct communication between any pair of processes.
This paper aims to examine the use of protocol complexes in the study of network computing. In this case, processes are located at the network nodes and communicate by exchanging messages only along the network's edges (i.e., not every pair of processes can directly communicate).

There are several reasons why applying the topological approach to network computing can be challenging, and a prominent one is that node identifiers yield protocol complexes whose sizes grow exponentially with the size of the underlying network.
However, many of the problems studied in this context are of local nature, and their definitions do not depend on the identifiers or the network size.
We leverage this independence to meet the above challenge and present \emph{local} protocol complexes, whose sizes do not depend on the network size.
As an application of the ``compacted'' protocol complexes, we reformulate the celebrated lower bound of $\Omega(\log^*n)$ rounds for 3-coloring the $n$-node ring in the topological framework.  
\end{abstract}

\thispagestyle{empty}
\newpage 
\setcounter{page}{1}

\section{Context and Objective}

Several techniques for formalizing distributed computing based on algebraic topology have emerged in the last decades, including the study of complexes capturing all possible global states of the systems at a given time~\cite{HerlihyKR2013}, and the study of the (di)homotopy classes of directed paths representing the execution traces of concurrent programs~\cite{FajstrupGHMR2016}. We refer to~\cite{GoubaultMT18} for a recent attempt to reconcile the two approaches. This paper is focusing on the first approach, based on the study of complexes.

\subparagraph*{Protocol Complexes.}~
A generic methodology for studying distributed computing through the lens of topology has been set by Herlihy and Shavit~\cite{HerlihyS99}. This methodology has played an important role in distributed computing, mostly for establishing impossibility results and time lower bounds~\cite{CastanedaR10,HerlihyS99,SaksZ00,FraigniaudPR22}, but also for establishing time upper bounds~\cite{CastanedaR12,AttiyaCHP19,HoestS06}. It is based on viewing  distributed computation as a topological deformation of an input space. More specifically, recall that a \emph{simplicial complex}~$\m{K}$ is a collection of non-empty subsets of a finite set~$V$, downward closed under inclusion, i.e., for every $\sigma\in \m{K}$, and every non-empty $\sigma'\subset\sigma$, it holds that $\sigma'\in \m{K}$. Every $\sigma\in \m{K}$ is called a \emph{simplex}, and every $v\in V$ is called a \emph{vertex}. For instance, an undirected graph $G=(V,E)$ with $E \subseteq {V \choose 2}$, can be viewed as the complex $\m{K}=\{\{v\}:\,v\in V\}\cup E$ on the set $V$ of vertices. A \emph{sub-complex} of a complex $\m{K}$ is a subset of simplices of $\m{K}$ forming a complex. 
The dimension of a simplex is one less than the number of its elements. A \emph{facet} of a complex $\m{K}$ is a maximal simplex of $\m{K}$, that is, a simplex  not contained in any other simplex. E.g., the facets of a graph are its edges and its isolated nodes (viewed as singleton sets). We note that a set of facets uniquely defines a complex.

The set of all possible input (resp., output) configurations of a distributed system can be viewed as a simplicial complex, called \emph{input complex} (resp., \emph{output complex}), and denoted by~$\m{I}$ (resp.,~$\m{O}$). A vertex of $\m{I}$ (resp., $\m{O}$) is a pair $(p,x)$ where $p$ is a process name, and $x$ is an input (resp., output) value. For instance, the input complex of binary consensus in an $n$-process system with process names $p_1,\dots,p_n$ is:
\[
\m{I}_{\mbox{\scriptsize $\|$}}=\Big\{\big\{(p_i,x_i): i\in I, x_i\in\{0,1\} \; \mbox{for every $i\in I$}\big\}: I\subseteq [n], I\neq \varnothing\Big\}, 
\]
with $[n]=\{1,\dots,n\}$, and the output complex is:
\[
\m{O}_{\mbox{\scriptsize $\|$}}=\Big\{\big\{(p_i,y): i\in I \big\}: I\subseteq [n], I\neq \varnothing, y \in\{0,1\}\Big\}. 
\]
One can check that $\m{I}_{\mbox{\scriptsize $\|$}}$ and $\m{O}_{\mbox{\scriptsize $\|$}}$ are indeed collections of non-empty subsets of a finite set, downward closed under inclusion. A distributed computing \emph{task} is then specified as a \emph{carrier map} $\Delta:\m{I}\to 2^{\m{O}}$, i.e., a function~$\Delta$ that maps every input simplex $\sigma\in\m{I}$ to a sub-complex $\Delta(\sigma)$ of the output complex, satisfying that, for every $\sigma,\sigma'\in \m{I}$, if $\sigma\subseteq\sigma'$ then $\Delta(\sigma)$ is a sub-complex of~$\Delta(\sigma')$. The carrier map~$\Delta$ is describing the output configurations that are legal with respect to the input configuration~$\sigma$. For instance, the specification of consensus is, for every $\sigma= \{(p_i,x_i): i\in I, x_i\in\{0,1\} \}\in\m{I}_{\mbox{\scriptsize $\|$}}$, 
\[
\Delta_{\mbox{\scriptsize $\|$}}(\sigma)=\left\{\begin{array}{ll}
\big\{\{(p_i,0):i\in I\},\; \{(p_i,1):i\in I\}\big \} & \mbox{if} \; \exists\, i,j\in I, \, x_i\neq x_j;\\
\big\{\{(p_i,y):i\in I\}\big \}  & \mbox{if} \; \forall \, i \in I, \, x_i=y.
\end{array}\right.
\]
Note that the specification of consensus given here is very general, i.e., $\Delta$ is specified for every simplex $\sigma\in\m{I}_{\mbox{\scriptsize $\|$}}$. This enables, e.g., to handle crash failures. In absence of failures, the specification of a task can be done just by specifying $\Delta$ for the facets in the input complex. 

In the topological framework, computation is modeled by a \emph{protocol complex} that evolves with time, where the notion of ``time'' depends on the computational model at hand. The protocol complex at time~$t$, denoted by $\m{P}^{(t)}$, captures all possible states of the system at time~$t$.  Typically, a vertex of $\m{P}^{(t)}$ is a pair $(p,s)$ where $p$ is a process name, and $s$ is a possible state of $p$ at time~$t$. A set $\{(p_i,s_i): i\in I\}$ of such vertices, for $\emptyset\neq I\subseteq [n]$, forms a simplex of $\m{P}^{(t)}$ if the states $s_i$, $i\in I$, are mutually compatible, that is, if $\{s_i : i\in I\}$ forms a possible global state for the processes in the set $\{p_i : i\in I\}$ at time~$t$. 

A crucial point is that an algorithm that outputs in time $t$ induces a mapping $\delta: \m{P}^{(t)}\to \m{O}$. Specifically, 
if the process~$p_i$ with state~$s_i$ at time $t$ outputs~$y_i$, then~$\delta$ maps the vertex $(p_i,s_i)\in \m{P}^{(t)}$ to the vertex $\delta(p_i,s_i)=(p_i,y_i)$ in~$\m{O}$. 
For the task to be correctly solved, the mapping~$\delta$ must preserve the simplices of $\m{P}^{(t)}$, and must agree with the specification~$\Delta$ of the task. That is, $\delta$~must map simplices to simplices, and if the configuration $\{(p_i,s_i), i\in I\}$ of a distributed system is reachable at time~$t$ starting from the input configuration $\{(p_i,x_i), i\in I\}$, then it must be the case that 
\[
\{\delta(p_i,s_i), i\in I\}\in \Delta(\{(p_i,x_i), i\in I\}).
\]
The set of configurations reachable in time $t$ stating from an input configuration $\sigma\in\m{I}$ is denoted by $\Xi_t(\sigma)$. In particular, $\Xi_t:\m{I}\to 2^{\m{P}^{(t)}}$ is a carrier map.  

\subparagraph*{Fundamental Lemma.}
%
The framework defined by Herlihy and Shavit~\cite{HerlihyS99} enables to characterize the power and limitation of distributed computing, thanks to the following generic result, which can be viewed as the basis of distributed computing within the topological framework. Let us consider some (deterministic) distributed computing model, assumed to be \emph{full information}, that is, every process communicates its entire history at each of its communication steps. The following result connects solvability of a task by an algorithm in a given model with the existence of a mapping of a specific form between the topological complexes corresponding to this task and this model (see~\cite{CastanedaFPRRT21,HerlihyKR2013,HerlihyS99,HerlihyR97,HerlihyRT98} for instantiations of this result for different computational models). 

\begin{lemma}
\label{lem:generic}
A task $(\m{I},\m{O},\Delta)$ is solvable in time~$t$ if and only if there exists a simplicial map $\delta:\m{P}^{(t)}\to \m{O}$ such that, for every $\sigma\in\m{I}$,  $\delta(\Xi_t(\sigma))\subseteq \Delta(\sigma)$.
\end{lemma}

\begin{figure}[htb]
	\centering
	\includegraphics[scale=0.28]{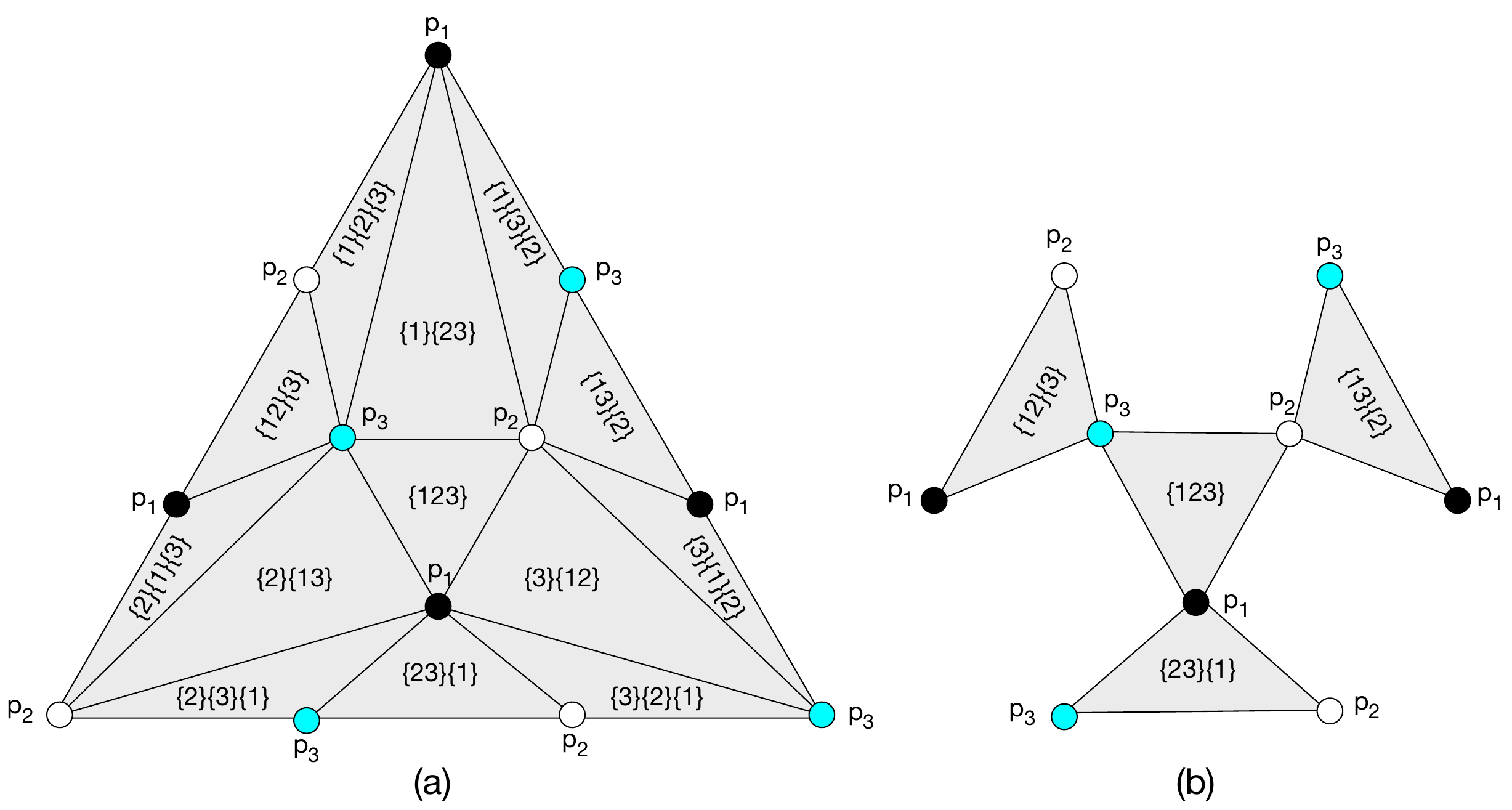}
	\caption{\sl (a) A chromatic subdivision of a 3-process simplex; (b)~Subdivision for 1-resiliency; a triangle labeled, e.g., $\{i\}\{jk\}$ corresponds to the case in which $p_i$ writes and reads the memory without seeing $p_j$ and $p_k$, while $p_j$ and $p_k$ saw $p_i$ when they read after they wrote, and they also saw each other; all possible interleavings for one write-read instruction are displayed.
	} 
	\label{fig:chromaticsubdivision}
\end{figure}

Again, beware that the notion of \emph{time} in the above theorem depends on the computational model.
The topology of the protocol complex~$\m{P}^{(t)}$, or, equivalently, the nature of the carrier map $\Xi_t$, depends on the input complex~$\m{I}$, and on the computing model at hand. 
For instance, wait-free computing in asynchronous shared memory systems induces protocol complexes by a deformation of the input complex, called \emph{chromatic subdivisions}~\cite{HerlihyKR2013} and depicted in Figure~\ref{fig:chromaticsubdivision}(a). 
Similarly, $t$-resilient computing may introduce \emph{holes} in the protocol complex, in addition to chromatic subdivisions, see Figure~\ref{fig:chromaticsubdivision}(b).  
More generally, the topological deformation $\Xi_t$ of the input complex caused by the execution of a full information protocol in the considered computing model entirely determines the existence of a decision map~$\delta:\m{P}^{(t)}\to\m{O}$, which makes the task $(\m{I},\m{O},\Delta)$ solvable or not in that model. 

	\subparagraph*{Topological Invariants.}
%
The typical approach for determining whether a task (e.g., consensus) is solvable in $t$ rounds
goes through identifying topological \emph{invariants}, i.e., properties of complexes that are preserved by simplicial maps. Specifically, the approach  consists in: 
\begin{enumerate}
\item Identifying a topological invariant, i.e., a property satisfied by the input complex~$\m{I}$, and preserved by~$\Xi_t$; 
\item Checking whether this invariant, which must be satisfied by the sub-complex $\delta(\m{P}^{(t)})$ of the output complex~$\m{O}$,  does not  contradict the specification $\Delta$ of the task. 
\end{enumerate}
For instance, in the case of binary consensus, the input complex $\m{I}_{\mbox{\scriptsize $\|$}}$ is a \emph{sphere}. One basic property of spheres is being \emph{path-connected} (i.e., there is a path in $\m{I}_{\mbox{\scriptsize $\|$}}$ between any two vertices). As mentioned earlier, shared-memory wait-free computing corresponds to subdividing the input complex~\cite{HerlihyKR2013}. Therefore, independently from the length~$t$ of the execution, the protocol complex $\m{P}^{(t)}$ is a chromatic subdivision of the  sphere $\m{I}_{\mbox{\scriptsize $\|$}}$, and thus it remains path-connected. 
On the other hand, the output complex $\m{O}_{\mbox{\scriptsize $\|$}}$ of binary consensus is the disjoint union of two complexes $\m{O}_0$ and~$\m{O}_1$, where $\m{O}_y=\big\{\{(i,y): i\in I \}, I\subseteq [n], I\neq \varnothing\big\}$ for $y\in\{0,1\}$. Since simplicial maps preserve connectivity, it follows that $\delta(\m{P}^{(t)})\subseteq \m{O}_0$ or $\delta(\m{P}^{(t)})\subseteq \m{O}_1$. 
As a consequence, $\delta$ cannot agree with $\Delta_{\mbox{\scriptsize $\|$}}$, as the latter maps the simplex $\{(i,0),i\in [n]\}$ to $\m{O}_0$, and the simplex $\{(i,1),i\in [n]\}$ to $\m{O}_1$. Therefore, consensus cannot be achieved wait-free, regardless of the number $t$ of rounds. 

The fact that connectivity plays a significant role in the inability to solve consensus in the presence of asynchrony and crash failures is known since the original proof of the FLP theorem~\cite{FischerLP85} in the early 1980s. However, the relation between $k$-set agreement and higher dimensional forms of  connectivity (i.e., the ability to contract high dimensional spheres) was only established ten years later~\cite{HerlihyS99,SaksZ00}. We refer to~\cite{HerlihyKR2013} for numerous applications of Lemma~\ref{lem:generic} to various models of  distributed computing, including asynchronous crash-prone shared-memory or fully-connected message passing models. In particular, for tasks such as renaming, identifying the minimal number~$t$ of rounds enabling a simplicial map~$\delta$ to exist is currently the only known  technique for upper bounding their time complexities~\cite{AttiyaCHP19}.

\begin{figure}[htb]
\centering
\includegraphics[scale=0.45]{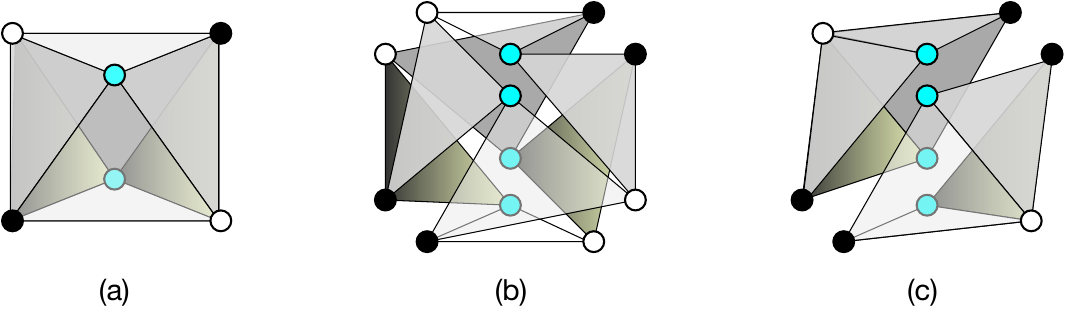}
\caption{\sl (a) The input complex of binary consensus for three processes; (b) The scissor cuts for the consistently directed 3-process cycle $C_3$ after one round; (c) The scissor cuts for the directed 3-process star~$S_3$, where edges are directed from the center to the leaves, after one round.} 
\label{fig:scissorcuts}
\end{figure}

	\subparagraph*{Network Computing.}
%
Recently, Casta\~neda et al.~\cite{CastanedaFPRRT21} applied Lemma~\ref{lem:generic}  to synchronous fault-free computing in networks, that is, to the framework in which processes are located at the vertices of a simple (no multiple edges, no loops) $n$-node undirected graph~$G$, and can exchange messages only along the edges of that graph. They mostly focus on \emph{input-output tasks} such as consensus and set-agreement, in a simplified computing model, called \textsf{KNOW-ALL}, specifying that every process is initially aware of the name and the location of all the other processes in the network. As observed in~\cite{CastanedaFPRRT21}, synchronous fault-free computing in the \textsf{KNOW-ALL} model preserves the \emph{facets} of the input complex, and does not subdivide them. 
However, \emph{scissor cuts} may occur between adjacent facets during the course of the computation, that is, the protocol complex $\m{P}^{(t)}$ is obtained from the input complex $\m{I}$ by partially separating facets that initially shared a simplex. Figure~\ref{fig:scissorcuts} illustrates two types of scissor cuts applied to the sphere, corresponding to two different communication networks. The positions of the cuts depend on the structure of the graph~$G$ in which the computation takes place, and determining the precise impact of the structure of $G$ on the topology of the protocol complex is a nontrivial challenge, even in the \textsf{KNOW-ALL} model. 

Instead, we aim at analyzing classical \emph{graph problems} (e.g., coloring, independent set, etc.) in the standard \local\/ model~\cite{Peleg01} of network computing, which is weaker than the \textsf{KNOW-ALL} model, and thus allows for more complicated topological deformations. In the \local\/ model, every node is initially aware of solely its identifier (which is unique in the network), and its input (e.g., for minimum weight vertex cover or for list-coloring), all nodes wake up synchronously, and compute in locksteps.  The \local\/ model is an ideal model for studying \emph{locality} in the context of network computing~\cite{Peleg01}. 

In addition to the fact that the topological deformations of the protocol complexes strongly depend on the structure of the network, another obstacle that makes applying the topological approach to the \local\/ model even more challenging is the presence of process \emph{identifiers}. Indeed, the model typically assumes that the node IDs are taken from a range $[N]$ where $N=\poly(n)$. As a consequence, independently from the potential presence of  other input values, the size of the complexes (i.e., their number of vertices) may become as large as ${N \choose n}n!$, since there are ${N \choose n}$ ways of choosing $n$ IDs, and $n!$ ways of assigning the $n$ chosen IDs to the $n$ nodes of $G$ (unless $G$ presents symmetries). For instance, Figure~\ref{fig:scissorcuts} assumes the \textsf{KNOW-ALL} model, hence fixed IDs. Redrawing these complexes assuming  that the three processes can pick arbitrary distinct IDs as in the \local\/ model, even in the small domain $\{1,2,3,4\}$, would yield a cumbersome figure with 24 nodes. Note that the presence of IDs also results in input complexes that may be topologically more complicated than pseudospheres, even for tasks such as consensus. 

Importantly, the fact that the IDs are not fixed a priori, and may even be taken from a range exceeding~$[n]$, is inherent to distributed network computing. Indeed, this framework aims at understanding the power and limitation of computing in large networks, from LANs to the whole Internet, where the processing nodes are  assigned arbitrary IDs taken from a range of values which may significantly exceed the number of nodes in the network. 
	\subparagraph{Objective.}
To sum up, while the study of protocol complexes has found numerous applications in the context of fault-tolerant message-passing or shared-memory computing, extending this theory to network computing faces a difficulty caused by the presence of arbitrary IDs, which are often the only inputs to the processes~\cite{Peleg01}. The objective of this paper is to show how the combinatorial blowup caused by the presence of IDs in network computing can be avoided, at least as far as local computing  is concerned. 

\section{Our Results}

We show how to bypass the aforementioned exponential blowup in the size of the complexes, that would result from a straightforward application of Lemma~\ref{lem:generic} for analyzing the complexity of tasks in networks. Our result holds for a variety of problems, including classical graph problems such as vertex and edge-coloring, maximal independent set (MIS), maximal matching, etc. More specifically, it holds for the large class of \emph{locally checkable labeling} (LCL) tasks~\cite{NaorS95} on bounded-degree graphs. These are tasks for which it is possible to locally verify the correctness of a solution, and thus they are sometimes viewed as the analog of NP in the context of computing in networks. An LCL task is described by a finite set of labels, and a local description of how these labels can be legally assigned to the nodes of a network. 
Our local characterization theorem is strongly based on a seminal result by Naor and Stockmeyer~\cite{NaorS95} who showed that the \emph{values} of the IDs do not actually matter much for solving LCL tasks in networks, but only their \emph{relative order} does. 

We prove an analog of Lemma~\ref{lem:generic}, but where the size of the complexes involved in the statement is independent of the size of the networks. Specifically, the size of the complexes in our characterization theorem depends solely on the maximum degree~$d$ (number of neighbors) in the network, the number of labels used for the description of the task, and the number of rounds (time) of the considered algorithm for solving that task. In particular, the identifiers are taken from a bounded-size set, even if the theorem applies to tasks defined on networks with arbitrarily large number~$n$ of nodes, and for identifiers taken from an arbitrarily large range $[N]$.
We denote by $\m{K}_{d,[R]}$ the fact that the facets of~$\m{K}$ have dimension~$d$, and that the IDs in it are taken from the set~$[R]=\{1,\dots,R\}$, and we let $\m{K}_{d}=\m{K}_{d,\varnothing}$.
In addition, $\pi: \m{K}_{d,[R]}\to \m{K}_{d}$ denotes the mapping that removes the IDs of the vertices. Every LCL task in networks with maximum degree~$d$ can be expressed topologically as a task $(\m{I}_d,\m{O}_d,\Delta)$ where $\m{I}_d$ and $\m{O}_d$ are complexes of dimension~$d$. Our main result is the following. 
\begin{theorem}[A simplified version of Theorem~\ref{theo:main}]
	\label{theo:main-simplified}
	For every LCL task $T=(\m{I}_d,\m{O}_d,\Delta)$ on graphs of maximum degree~$d$, and for every $t\geq 0$, there exists $R\in\mathbb{N}$ such that the following holds. 
	The task $T$ is solvable in $t$ rounds in the \local\/ model if and only if there is a simplicial map $\delta:\m{P}_{d,[R]}^{(t)}\to \m{O}_{d}$ such that, for every facet $\sigma\in\m{I}_{d,[R]}$,  $\delta(\Xi_t(\sigma))\subseteq \Delta(\pi(\sigma))$. 
\end{theorem}

Figure~\ref{fig:commdiag} provides a rough description of the commutative diagram corresponding to the brute force application of Lemma~\ref{lem:generic} to LCL tasks, and of the commutative diagram corresponding to Theorem~\ref{theo:main-simplified}. 
Note that Lemma~\ref{lem:generic}, which corresponds to the left diagram in Figure~\ref{fig:commdiag}, involves \emph{global} complexes with $(n-1)$-dimensional facets, whose vertices are labeled by IDs in an arbitrarily large set~$[N]$. In contrast, the complexes corresponding to Theorem~\ref{theo:main-simplified}, which correspond to the right diagram, are \emph{local} complexes, with facets of constant dimension, and  vertices labeled with IDs in a finite set whose size is constant w.r.t.\ the number of nodes~$n$ in the network. 

\begin{figure}[htb]
\begin{center}
\small
\begin{tikzpicture}
\matrix (m) 
[matrix of math nodes,row sep=4em,column sep=4em,minimum width=2em] 
{
\m{I}_{n-1,[N]} &  \m{P}_{n-1,[N]}^{(t)} \\
          &  \m{O}_{n-1,[N]}  \\ 
}; 
\path[-stealth] 
(m-1-1) edge node [above] {$\Xi_t$} (m-1-2)
        edge node [below] {$\Delta$} (m-2-2)
(m-1-2) edge node [right] {$\delta$} (m-2-2);
\end{tikzpicture}
\hspace{3cm}
\begin{tikzpicture}
\matrix (m) 
[matrix of math nodes,row sep=4em,column sep=4em,minimum width=2em] 
{
\m{I}_{d,[R]} &  \m{P}_{d,[R]}^{(t)} \\
 \m{I}_{d}       &  \m{O}_{d}  \\ 
}; 
\path[-stealth] 
(m-1-1) edge node [above] {$\Xi_t$} (m-1-2)
 (m-2-1)        edge node [above] {$\Delta$} (m-2-2)
(m-1-1) edge node [left] {$\pi$} (m-2-1)
(m-1-2) edge node [left] {$\delta$} (m-2-2);
\end{tikzpicture}
\end{center}
\vspace*{-5ex}
\caption{\sl The commutative diagrams of Lemma~\ref{lem:generic} (left), and Theorem~\ref{theo:main-simplified} (right). }
\label{fig:commdiag}
\end{figure}

As an application of Theorem~\ref{theo:main-simplified}, we reformulate the celebrated $\Omega(\log^\ast n)$ lower bound  rounds for 3-coloring the $n$-node ring-shaped network by Linial~\cite{Linial92}, in the algebraic topology framework (see Corollary~\ref{cor:reproving-linial}). 

Reducing the size of the protocol complex (and the other simplicial complexes involved) is standard in the highly studied case of \emph{colorless tasks}~\cite{BorowskyGLR01,HerlihyR12}.
This is a class of tasks where processes can adopt each other's input and output values, such as consensus, set agreement and approximate agreement. 
However, we stress that in our context of network computing and LCL tasks, almost all interesting tasks are not colorless,
which requires the use of another tool --- local complexes. 

\section{Models and Definitions}

We study networks modeled by simple, undirected $n$-node graphs, denoted $G=(V,E)$.
The \emph{degree} of a node is the number of its neighbors, and we are particularly interested in \emph{$d$-regular} graphs, where each node has degree~$d$.
A graph is connected if there is a path between every two nodes in it.
For $d\geq 2$, we denote by $\m{G}_d$ be the class of connected simple undirected $d$-regular graphs.
A \emph{star} is a graph composed of a center node that is a neighbor of all other nodes, and no additional edge; it can also be seen as a rooted tree of depth 1.
Given an $n$-node graph $G$, we study the collection of $n$ stars defined by each node and its neighbors.

Our study cases involve graph problems, where each node must be assigned a label satisfying specific conditions.
A proper \emph{$c$-coloring} of a graph is a function $\lambda:V\to\{1,\ldots,c\}$ such that for every pair of adjacent nodes $u,v$, it holds that $\lambda(u)\neq\lambda(v)$.
In the distributed setting, we want each node~$u$ to compute its color~$\lambda(u)$,
so that the resulting coloring is proper.

An \emph{independent set} is a set $S\subseteq V$ of node such that no two nodes in $S$ are neighbors.
Such an independent set is \emph{maximal} if no node outside of $S$ can be added to it without violating the independence condition.
An independent set $S$ can be represented by its indicator function $\lambda:V\to\{0,1\}$,
where $\lambda(u)=1$ if and only if $u\in S$.
In the distributed setting, \emph{maximal independent set} (MIS) is the task of assigning each node a Boolean value such that the set of all nodes assigned $1$ forms a maximal independent set.

\subsection{The \local\/ model}
The \local\/ model was introduced more than a quarter of a century ago (see, e.g., \cite{Linial92,NaorS95,Peleg01}) for studying which tasks can be solved \emph{locally} in networks, that is, which tasks can be solved when every node is bounded to collect information only from nodes in its vicinity.
Specifically, the  \local\/ model states that the  processors are located at the nodes of a connected simple graph $G=(V,E)$ modeling a network. All nodes are fault-free, they wake up simultaneously, and they execute the same algorithm. Computation proceeds in synchronous rounds, where a round consists of the following three steps performed by every node: (1)~sending a message to each neighbor in~$G$, (2)~receiving the messages sent by the neighbors, and (3)~performing local computation. There are no bounds on the size of the messages exchanged at every round between neighbors, and there are no limits on the individual computational power or memory of the nodes. These assumptions enable the design of unconditional  lower bounds on the number of rounds required for performing some task (e.g., for providing the nodes with a proper coloring), while the vast majority of the algorithms solving these tasks do not abuse of these assumptions~\cite{Suomela13}, that is, they exchange small (i.e., polylogarithmic size) messages, and perform efficient (i.e., poly-time) individual computations.  

Every node in the network has an identifier (ID) which is supposed to be unique in the network. In $n$-node networks, the IDs are supposed to be in a range $1,\dots,N$ where $N\gg n$ typically holds (most often, $N=\poly(n)$). 
The absence of limits on the amount of communication and computation that can be performed at every round implies that the  \local\/ model enables \emph{full-information} protocols, that is, protocols in which, at every round, every node sends  all the information it acquired during the previous rounds to its neighbors. Therefore, for every $t\geq 0$, and every graph $G$, a $t$-round algorithm allows every node in~$G$ to acquire a local \emph{view} of~$G$, which is a ball of radius~$t$ in~$G$ centered at that node. 
A view includes the inputs and the IDs of the nodes in the corresponding ball.
It follows that a $t$-round algorithm  in  the  \local\/ model can be considered as a function from the set of views of radius~$t$ to the set of output values. 

	\subsection{Locally Checkable Labelings (LCL)}
A \emph{locally checkable labeling} (LCL)~\cite{NaorS95} 
is a graph problem on regular graphs that can be defined using a set $\m{L}$ of node-labels, and a set of labeled stars called \emph{good} stars.
For $d\geq 2$, an LCL for $d$-regular graphs involves labeling the nodes of a graph $G=(V,E)\in\m{G}_d$ with a labeling $\lambda: V \to \m{L}$ such that every star in $G$ (defined by a node $v\in V$ and its neighbors) is assigned labels by $\lambda$ in a way that forms a good star.

For example,
a proper $c$-coloring in $\m{G}_d$ can be described by the labels $\{1,\dots,c\}$ and the collection of good stars where the center node has a color different from the colors of the leaves.
Similarly, a maximal independent set (MIS) in $\m{G}_d$ can be described by the label set $\{0,1\}$ and the collection of degree-$d$ stars where if the center node is labeled~$1$ then all the leaves are labeled~$0$ (independence), and if the center node is labeled~$0$ then at least one leaf is labeled~$1$ (maximality). 
Other tasks such as variants of coloring, or $(2,1)$-ruling set\footnote{Recall that an $(\alpha,\beta)$-ruling set in a graph $G=(V,E)$ is a set $R\subseteq V$ such that, for any node $v\in V$ there is a node $u\in R$ in distance at most~$\beta$ from~$v$, and any two nodes in $R$ are at distance at least~$\alpha$ from each other.} 
can be described similarly, by a finite number of properly labeled stars. 
  
Formally, given a finite set $\m{L}$ of labels, we denote by $\mathbf{S}_d^{\m{L}}$ the set of all labeled stars resulting from labeling each node of the $(d+1)$-node star by some label in~$\m{L}$. An LCL 
is then defined by a finite set $\m{L}$ of labels, and a set $\m{S}\subseteq \mathbf{S}_d^{\m{L}}$ of good stars; the stars in $\mathbf{S}_{d}^{\m{L}}\setminus \m{S}$ are called \emph{bad}. 
The computational task defined by an LCL $(\m{L},\m{S})$ consists, for every node of every graph $G\in \m{G}_d$, of computing a label in~$\m{L}$ for each node in $G$ such that each resulting labeled star in $G$ is isomorphic to a star in~$\m{S}$. In other words, the objective of every node is to compute a label in $\m{L}$ such that every resulting labeled star in $G$ is good. It is undecidable, in general, whether a given LCL task has an algorithm performing in $O(1)$ rounds in the \local\/ model~\cite{NaorS95}. 

More generally, LCL tasks include tasks in which nodes have inputs, potentially of some restricted format. 
For instance, this is the case of the task consisting of reducing $c$-coloring  to MIS in the $n$-node cycle $C_n$, studied in the next section. 
In this case, an LCL task is described by a quadruple $(\m{L}_{in},\m{S}_{in}, \m{L}_{out},\m{S}_{out})$ where $\m{L}_{in}$ and $\m{L}_{out}$ are the input and output labels, respectively. The set of stars $\m{S}_{in}$ can often be simply viewed as a promise stating that every star of the input graph $G$ belongs to~$\m{S}_{in}$, and the set $\m{S}_{out}$ is  the target set of good stars. 

In its full generality, the framework of LCL tasks can  be extended by replacing stars by balls of radius $t>1$, for capturing more problems, like $(\alpha,\beta)$-ruling set for large $\alpha$'s or $\beta$'s. They can also be extended to non-regular graphs with bounded maximum degree~$d$. However, up to extending the set of labels, all such tasks can be reformulated in the context of stars and regular graphs~\cite{Brandt19}.
To get the intuition of why this is true, consider the task in which every node must compute a label in~$\{T,F\}$ such that every node labeled~$F$ has a node labeled~$T$ at distance at most~$k$, for some fixed $k\geq 1$. 
To describe this task by stars, let $\m{L}=\{T,F_1,\dots,F_k\}$, where we interpret the index~$i$ of a label $F_i$ as an upper bound on the distance to a $T$-marked node. The good stars are defined as follows: a star whose center is labeled~$T$ is always good, and, for $i=1,\dots,k$, a star whose center is labeled~$F_i$ is good if it has a leaf with label in $\{T,F_1,\dots,F_{i-1}\}$. 

In another, more general case of LCLs, the legality of an output star may depend on the corresponding input star~\cite{NaorS95}. 
In this scenario, an LCL is defined by a quintuple (a 5-tuple), consisting of input labels and stars, output labels and stars, and a relation between the input and output stars. A typical example of such a setting is list-coloring, where the output color of each node must be chosen from a list of colors provided as input to the node. 
To simplify the presentation, we consider LCL tasks without an input-output relation and stick to the quadruple representation. Nevertheless, handling LCLs with input-output relations is a simple extension of our techniques, and we explain how to apply it after presenting the topological definition of LCLs, as defined in  Definition~\ref{def:IandOforLCL}.

\section{Warm Up: Coloring and MIS in the Ring}
\label{sec:warmup}

In this section, we exemplify our technique, in a way that resembles the proof of Theorem~\ref{theo:main-simplified}. 
We consider an LCL task on a ring, where the good input stars define a proper $3$-coloring, and the good output stars define a maximal independent set (MIS). That is, we study the time complexity of reducing a $3$-coloring to a MIS on a ring. It is known~\cite{Linial92}  that there is a 2-round algorithm for the problem in the \local\/ model, and we show that this is optimal using topological arguments.
This toy example provides the basic concepts and arguments that  we use later, when considering general LCL tasks and proving Theorem~\ref{theo:main-simplified}. 

\subsection{Reduction from 3-coloring to MIS}
\label{subsec:reduction3colMIS}

\begin{figure}[!h]
\centering
\includegraphics[scale=0.4]{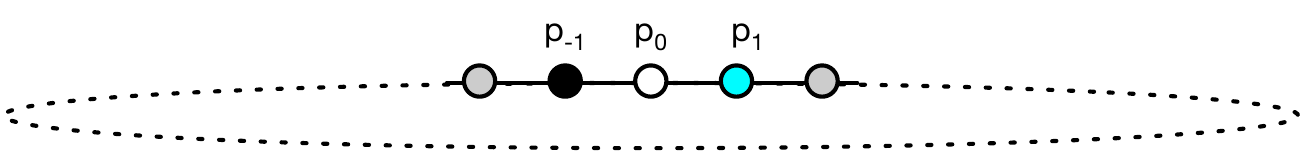}
\caption{\sl Three consecutive nodes in the $n$-node ring.}
\label{fig:ring3nodes}
\end{figure}

Let us consider three consecutive nodes of the $n$-node ring $C_n$, denoted by $p_{-1},p_0$, and $p_1$, as displayed on Figure~\ref{fig:ring3nodes}. 
Note that the names $p_{-1},p_0$ and $p_1$ are arbitrary, and external to the algorithm.
Here and later, $p_0$ will always denote the central node in the star we analyze.

\begin{figure}[!h]
\centering
\includegraphics[scale=0.35]{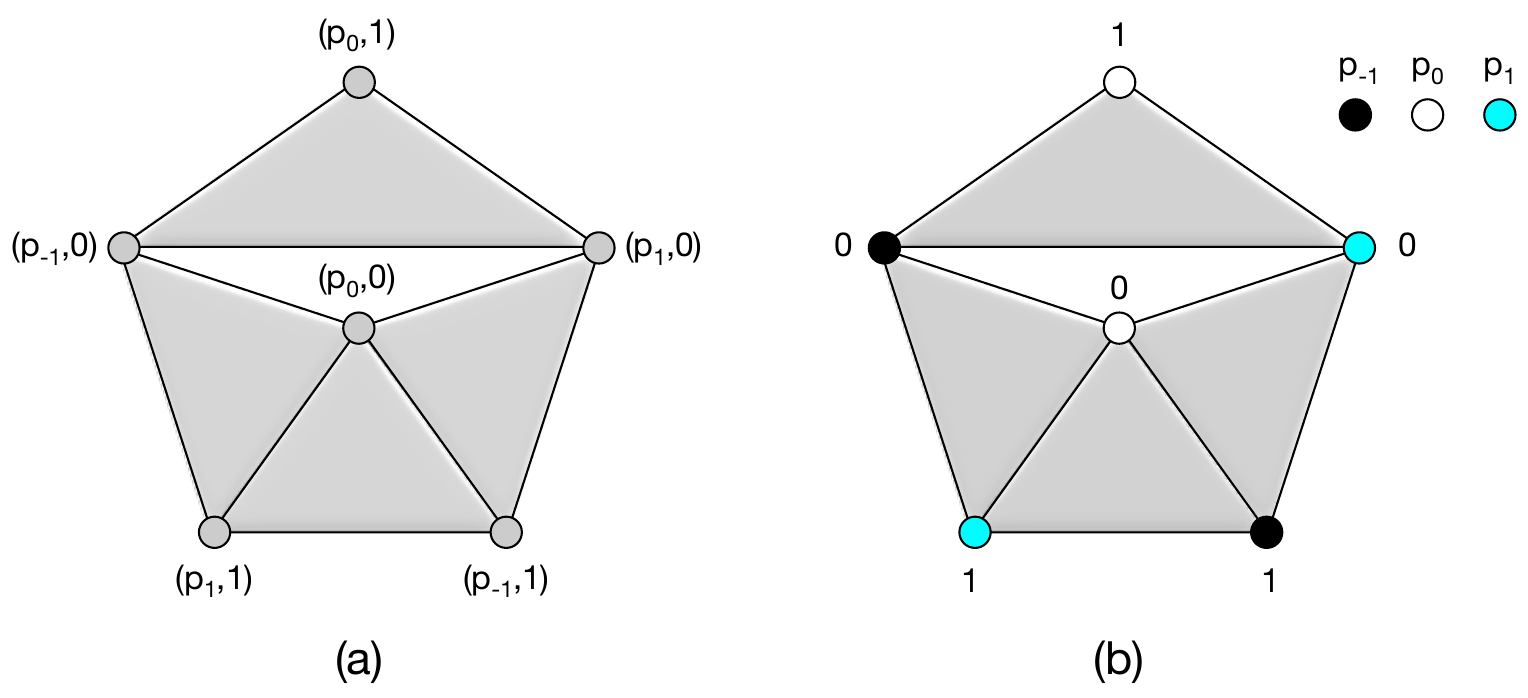}
\caption{\sl The local complex $\LMIS$ of MIS in the ring.
(a) the vertices are labeled with the index of the processes and the values; (b) the indexes of the processes are replaced by colors. 
}
\label{fig:complexMIS}
\end{figure}

We now apply topological tools in order to analyze this task.
By the independence property, if $p_0$ is in the MIS, then neither $p_{-1}$ nor $p_1$ can be in the MIS, and, by the maximality property, if  $p_0$ is not in the MIS, then $p_{-1}$ or $p_1$, or both, must be in the MIS. These constraints are captured by the complex $\LMIS$ displayed on Figure~\ref{fig:complexMIS}, including six vertices $(p_i,x)$, with $i\in \{-1,0,1\}$, and $x\in\{0,1\}$, where $x=1$ (resp., $x=0$) indicates that $p_i$ is in the MIS (resp., not in the MIS). 

The complex $\LMIS$ of Figure~\ref{fig:complexMIS} has four facets of dimension~2: they are triangles. Some triangles intersect along an edge, while some others intersect only at a node. The complex $\LMIS$ is called the \emph{local} complex of MIS in the ring (the index~2 refers to the fact that rings have degree~2). Note that the sets $\{(p_{-1},0),(p_0,0),(p_1,0)\}$ and $\{(p_{-1},1),(p_0,1),(p_1,1)\}$ do not form simplices of $\LMIS$. 
We call these two sets monochromatic. 
 In the objective of reducing 3-coloring to MIS,  $\LMIS$ will be the output complex, corresponding to $\m{O}_{d}$ with $d=2$ in Figure~\ref{fig:commdiag} and in Theorem~\ref{theo:main-simplified}.

\bigskip

\begin{figure}[htb]
\centering
\includegraphics[scale=0.35]{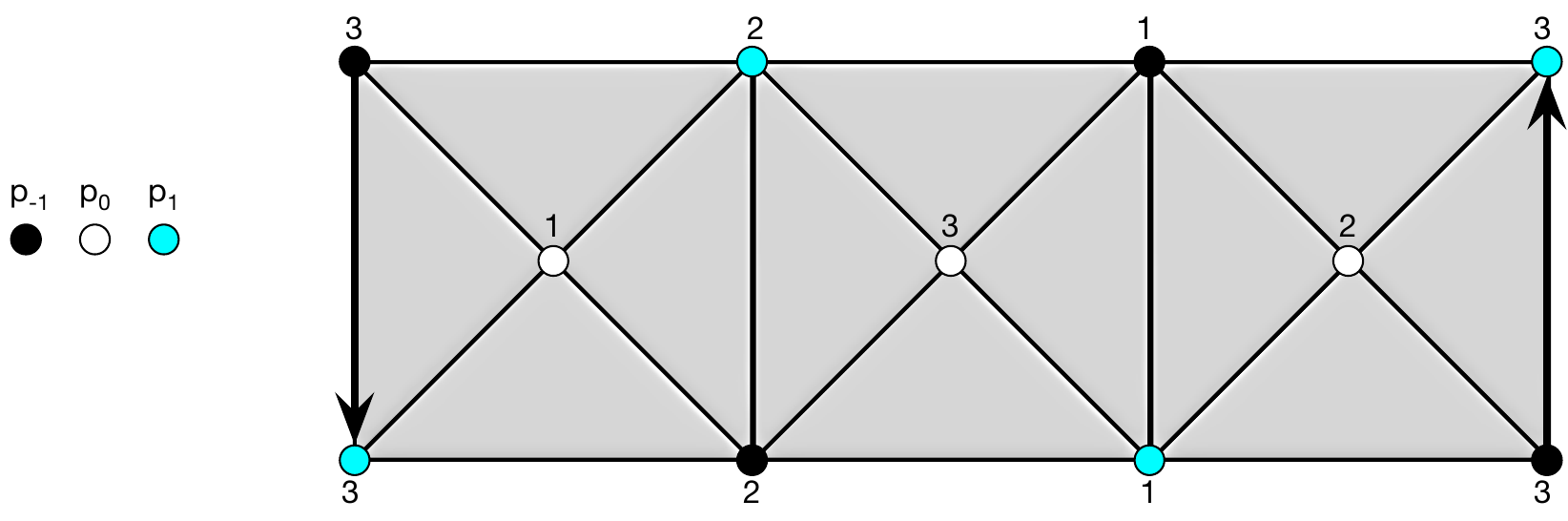}
\caption{\sl Local complex $\LCOL_2$ of 3-coloring in the ring.}
\label{fig:complex3coloring}
\end{figure}

Similarly, let us focus on 3-coloring, with the same three processes $p_{-1}, p_0$, and $p_1$.
The neighborhood of $p_0$ cannot include the same color as its own color, and thus there are twelve possible colorings of the nodes in the star centered at~$p_0$. Each of these stars corresponds to a 2-dimensional simplex, forming the facets of the local complex $\LCOL_2$ of 3-coloring in the ring, depicted in Figure~\ref{fig:complex3coloring}. This complex contains nine vertices of the form $(p_i,c)$, with $i\in \{-1,0,1\}$, and $c\in\{1,2,3\}$, and twelve facets. 
Note that the vertices $(p_{-1},3)$ and $(p_1,3)$ appear twice in the figure, since the leftmost and rightmost edges are identified, but in opposite direction, forming a M\"obius strip. $\LCOL_2$ is a manifold (with boundary). 
When reducing 3-coloring to MIS,  $\LCOL_2$ will be the input complex, corresponding to $\m{I}_{d}$ with $d=2$ in Figure~\ref{fig:commdiag}. 

\subparagraph{Remark.} 
It is crucial to note that the complexes displayed in figures~\ref{fig:complexMIS} and~\ref{fig:complex3coloring} are not the ones used in the standard settings (e.g.,~\cite{CastanedaFPRRT21,HerlihyKR2013}), for which Lemma~\ref{lem:generic} would use vertices of the form $(p,x)$ for $p\in[n]$, or even $p\in [N]$ assuming IDs in a range of $N$ values. 
As a consequence, these complexes have 6~vertices instead of $2n!{N \choose n}$ for MIS, and 9~vertices instead of $3n!{N \choose n}$ for coloring, where $n$ can be arbitrarily large. Even if the IDs would have been fixed, the approach of Lemma~\ref{lem:generic} would yield complexes with a number of vertices linear in~$n$, while the complexes of figs.~\ref{fig:complexMIS} and~\ref{fig:complex3coloring} are of constant sizes. 

\medskip

As it is well-know since the early work by Linial~\cite{Linial92}, a properly 3-colored ring can be ``recolored'' into a MIS in just two rounds. First, the nodes colored~3 recolor themselves~1 if they have no neighbors originally colored~1. Then, the nodes colored~2 do the same, i.e., they recolor themselves~1 if they have no neighbors colored~1 (whether it be neighbors originally colored~1, or nodes that recolored themselves~1 during the first round). The nodes colored~1 output~1, and the other nodes output~0. The set of nodes colored~1 forms a MIS. Note that this algorithm is \emph{name-independent}, i.e., it can run in an anonymous network. 

	\subparagraph{Task specification.} 
%
The specification of reducing 3-coloring to MIS can be given by the trivial carrier map 
$
\Delta:\LCOL_2\to 2^{\LMIS}
$ 
defined by 
$
\Delta(F)=\{F': F' \; \mbox{is a facet of} \; \LMIS\}
$
for every facet $F$ of~$\LCOL_2$. 
(As the \local\/ model is failure-free, it is enough to describe all maps at the level of facets.)
Note that the initial coloring of a facet in $\LCOL_2$ does not induce constraints on the facet of $\LMIS$ to which it should be mapped. 
 Figure~\ref{fig:ExampleRing} displays some of the various commutative diagrams that will be considered in this section. In all of them, $\Delta$ is the carrier map specifying reduction from 3-coloring to MIS in the ring, and none of the simplicial maps $\delta$ exist. Also recall that~$\pi$ is the map removing IDs. 
 
 \begin{figure}[htb]
\begin{center}
\small
\begin{tikzpicture}
\matrix (m) 
[matrix of math nodes,row sep=4em,column sep=4em,minimum width=2em] 
{
\LCOL_{2,\varnothing} &  \LCOL_{2,\varnothing} \\
 \LCOL_2       &  \LMIS  \\ 
}; 
\path[-stealth] 
(m-1-1) edge node [above] {$\Xi_0$} (m-1-2)
 (m-2-1)        edge node [above] {$\Delta$} (m-2-2)
(m-1-1) edge node [left] {$\pi$} (m-2-1)
(m-1-2) edge node [left] {$\delta$} (m-2-2);
\end{tikzpicture}
\begin{tikzpicture}
\matrix (m) 
[matrix of math nodes,row sep=4em,column sep=4em,minimum width=2em] 
{
\LCOL_{2,\varnothing} &  \m{P}_{2,\varnothing}^{(1)}  \\
 \LCOL_2       &  \LMIS  \\ 
}; 
\path[-stealth] 
(m-1-1) edge node [above] {$\Xi_1$} (m-1-2)
 (m-2-1)        edge node [above] {$\Delta$} (m-2-2)
(m-1-1) edge node [left] {$\pi$} (m-2-1)
(m-1-2) edge node [left] {$\delta$} (m-2-2);
\end{tikzpicture}
\begin{tikzpicture}
\matrix (m) 
[matrix of math nodes,row sep=4em,column sep=4em,minimum width=2em] 
{
 \LCOL_{2,[24]}  &   \LCOL_{2,[24]}  \\
 \LCOL_2       &  \LMIS \\ 
}; 
\path[-stealth] 
(m-1-1) edge node [above] {$\Xi_0$} (m-1-2)
 (m-2-1)        edge node [above] {$\Delta$} (m-2-2)
(m-1-1) edge node [left] {$\pi$} (m-2-1)
(m-1-2) edge node [left] {$\delta$} (m-2-2);
\end{tikzpicture}
\begin{tikzpicture}
\matrix (m) 
[matrix of math nodes,row sep=4em,column sep=4em,minimum width=2em] 
{
 \LCOL_{2,[R]}  &  \m{P}_{2,[R]}^{(1)} \\
 \LCOL_2       &  \LMIS \\ 
}; 
\path[-stealth] 
(m-1-1) edge node [above] {$\Xi_1$} (m-1-2)
 (m-2-1)        edge node [above] {$\Delta$} (m-2-2)
(m-1-1) edge node [left] {$\pi$} (m-2-1)
(m-1-2) edge node [left] {$\delta$} (m-2-2);
\end{tikzpicture}
\end{center}
\vspace*{-2ex}
\caption{\sl Complexes corresponding to reduction from 3-coloring to MIS in the $n$-node ring. 
From left to right: 0~rounds without IDs, 1-round without IDs, 0~rounds with ID, and 1-round with IDs.}
\label{fig:ExampleRing}
\end{figure}

\subsection{Name-Independent Algorithms}

We start by considering  \emph{name-independent} algorithms, i.e., algorithms where all nodes run the same algorithms and do not use their IDs. 
These algorithms can also be used in anonymous networks, where IDs do not exist.

\subsubsection{Impossibility in Zero Rounds}

\noindent\emph{Name-preserving maps.} 
Let us consider an alleged name-independent algorithm $\alg$ which reduces $3$-coloring to MIS in zero rounds.
Such an algorithm sees only the node's color $c\in\{1,2,3\}$, and must map it to some $x\in\{0,1\}$.
This induces a mapping $\delta$, that maps every pair $(p_i,c)$ with $i\in \{-1,0,1\}$ and $c\in\{1,2,3\}$, to a pair $\delta(p_i,c)=(p_i,x)$ with $x\in\{0,1\}$. 
We say that such a mapping is \emph{name-preserving},
i.e., the algorithm maps the vertices in Figure~\ref{fig:complex3coloring} to the vertices in  Figure~\ref{fig:complexMIS}(b) while preserving the names $p_{-1},p_0,p_1$ of these vertices. 
Therefore, the algorithm induces a \emph{name-preserving} simplicial map  $\delta:\LCOL_{2}\to\LMIS$.
The term name-preserving (sometimes refered to as \emph{chromatic}) is the formal way to express the fact that a vertex $(p,x)$ is mapped to a vertex $(p,y)$, that is, the name~$p$ is preserved.

As discussed above, we are interested in name-independent algorithms. 
In topological terms, such algorithms translate to name-preserving name-independent simplicial maps (we slightly abuse notation by using the terms name-preserving and  name-independent both for an algorithm and for a mapping).
We are therefore questioning the existence of a name-preserving name-independent simplicial map~$\delta:\LCOL_2\to\LMIS$. 
This is in correspondence with Figure~\ref{fig:commdiag} and Theorem~\ref{theo:main-simplified}, in the degenerate case where $t=0$ and $[R]=\varnothing$, for which $\LCOL_2=\m{I}_2$, and $\LCOL_{2,\varnothing}=\m{I}_{2,\varnothing}=\m{P}^{(0)}_{2,\varnothing}=\LCOL_2$ --- see the leftmost diagram in Figure~\ref{fig:ExampleRing}.

It is easy to see that there cannot exist a name-preserving name-independent simplicial map~$\delta$ from the manifold $\LCOL_2$ to $\LMIS$  (from Figure~\ref{fig:complex3coloring} to Figure~\ref{fig:complexMIS}(b)). Indeed, a simplicial map $\delta:\LCOL_2\to\LMIS$ can only map $\LCOL_2$ entirely to the sub-complex of $\LMIS$ induced by the simplex $\sigma_{00}=\{(p_{-1},0),(p_0,1),(p_1,0)\}$, or entirely to the sub-complex of $\LMIS$ induced by all the other simplices. To see why, assume the opposite. Then, w.l.o.g., we can assume that the vertex $(p_0,1)$ of $\LCOL_2$ is mapped to $(p_0,0)$ of $\LMIS$, and that $(p_0,3)$ of $\LCOL_2$ is mapped to $(p_0,1)$ of $\LMIS$. Let us consider the two simplices 
\[
\{(p_{-1},2),(p_0,1),(p_1,2)\} \; \mbox{and} \; \{(p_{-1},2),(p_0,3),(p_1,2)\}
\]
of $\LCOL_2$, which form a sub-complex of $\LCOL_2$. In order to preserve the edges of this sub-complex, $(p_{-1},2)$ and $(p_1,2)$ must be respectively mapped to $(p_{-1},0)$ and $(p_1,0)$. It follows that the simplex $\{(p_{-1},2),(p_0,3),(p_1,2)\}$ of $\LCOL_2$ is correctly mapped to a simplex of $\LMIS$ (specifically, to the simplex $\{(p_{-1},0),(p_0,1),(p_1,0)\}$). However, the simplex $\{(p_{-1},2),(p_0,1),(p_1,2)\}$ of $\LCOL_2$ is mapped to the monochromatic set $\{(p_{-1},0),(p_0,0),(p_1,0)\}$ which is not a simplex of $\LMIS$  (it is a hole in this complex as depicted in Figure~\ref{fig:complexMIS}), contradiction. Thus, $\LCOL_2$ must be entirely mapped to the sub-complex of $\LMIS$ induced by the simplex $\sigma_{00}=\{(p_{-1},0),(p_0,1),(p_1,0)\}$, or entirely to the sub-complex of $\LMIS$ induced by all the other simplices. This yields two cases: \\
~$-$ In the former case, $p_0$ outputs~1 independently from its input color, and therefore, by the name independence, $p_{-1}$ and $p_1$ also output~1, which is not the case in~$\sigma_{00}$.\\
~$-$  In the latter case, $p_0$ outputs~0 independently from its input color, and therefore, by the name independence, $p_{-1}$ and $p_1$ also output~0, yielding a contradiction as no monochromatic sets are simplices of $\LMIS$. \\
Hence, there are no name-preserving name-independent simplicial maps~$\delta:\LCOL_2\to\LMIS$. 
The absence of a name-preserving name-independent simplicial map $\delta:\LCOL_2\to\LMIS$ is a witness of the impossibility to construct a MIS from a 3-coloring of the ring in zero rounds, when using a name-independent algorithm.

\subsubsection{Impossibility in One Round}
\label{subsubsec:no1roundwithoutID}
\begin{figure}[htb]
\centering
\includegraphics[scale=0.25]{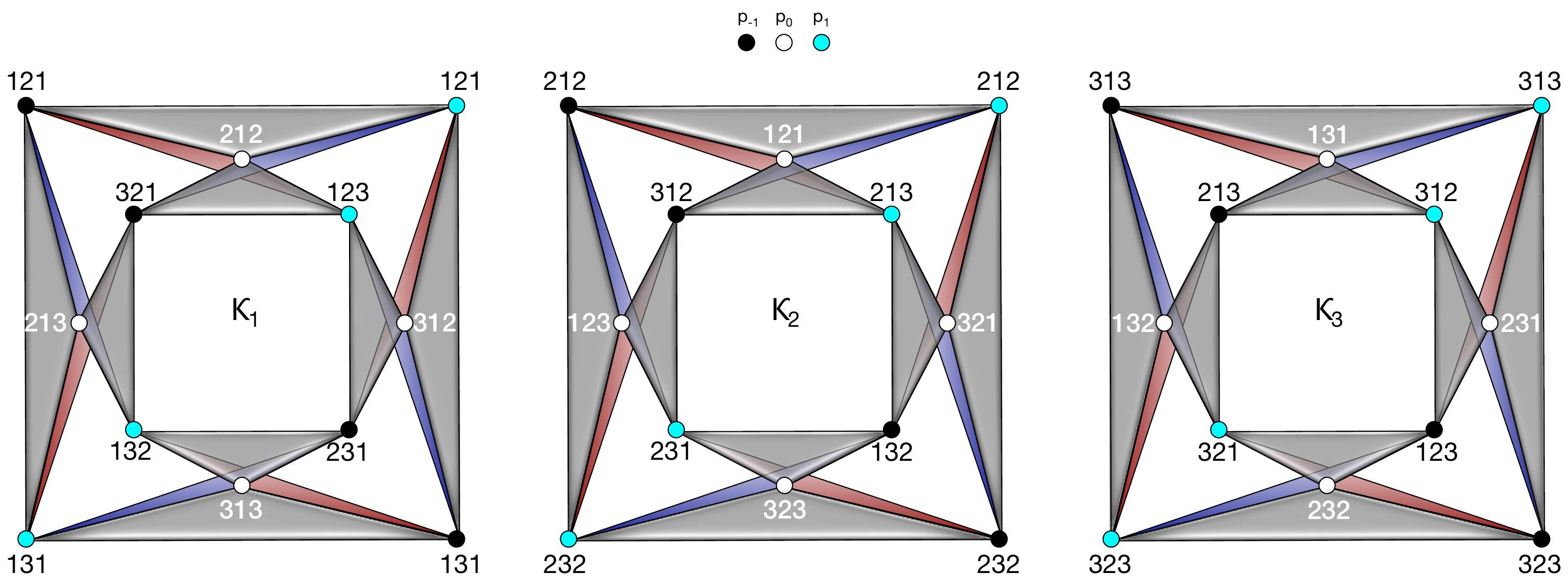}
\caption{\sl Local protocol complex $\m{P}_{2,\varnothing}^{(1)}$ after 1 round starting from a 3-coloring of the ring.}
\label{fig:complex3coloring1round}
\end{figure}

\begin{figure}[htb]
\centering
\includegraphics[scale=0.25]{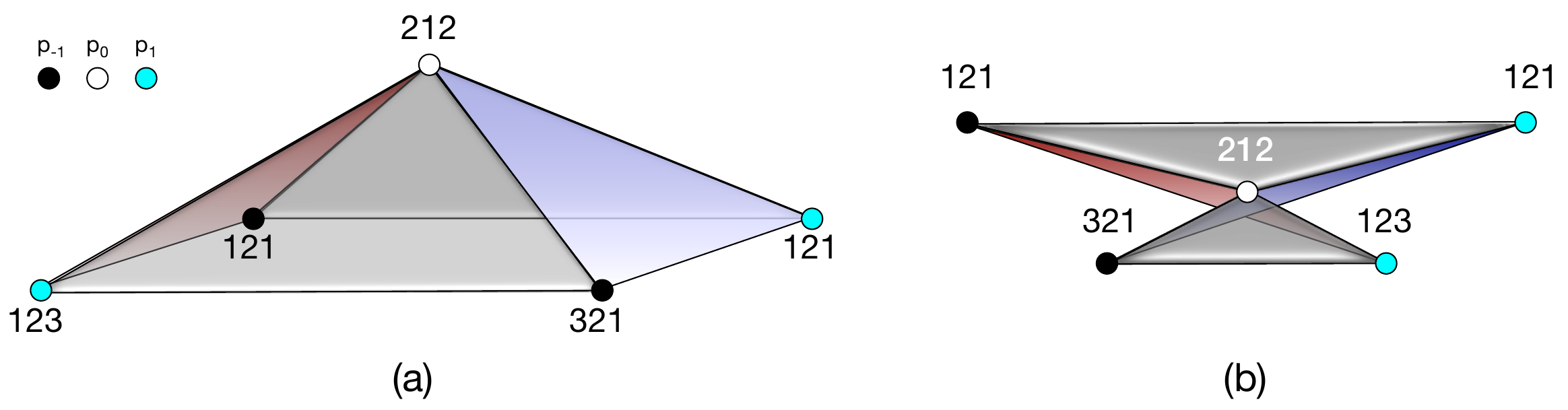}
\caption{\sl (a) A cone composed of four triangles; (b) The same cone ``twisted''. }
\label{fig:cone}
\end{figure}

For analyzing 1-round algorithms, let us consider the local protocol complex $\m{P}_{2,\varnothing}^{(1)}$, including the views of the three nodes $p_{-1}, p_0$, and $p_1$ after one round. The vertices of  $\m{P}_{2,\varnothing}^{(1)}$ are of the form $(p_i,xyz)$ with $i\in \{-1,0,1\}$, and $x,y,z\in\{1,2,3\}$, $x\neq y$, and $y\neq z$. The vertex $(p_i,xyz)$ is representing a process $p_i$ starting with color $y$, and receiving the input colors $x$ and $z$ from its left and right neighbors, respectively. The facets of $\m{P}_{2,\varnothing}^{(1)}$ are of the form $\{(p_{-1},x'xy),(p_0,xyz),(p_1,yzz')\}$. Figure~\ref{fig:complex3coloring1round} displays that complex, which consists of three connected components $\m{K}_1,\m{K}_2$, and $\m{K}_3$ where, for $y=1,2,3$, $\m{K}_y$ includes the four vertices $(p_0,xyz)$ for $x,z\in\{1,2,3\}\smallsetminus\{y\}$, and all triangles that include these vertices. Each set of four triangles sharing a vertex $(p_0,xyz)$ forms a cone (see Figure~\ref{fig:cone}). These cones are displayed twisted on Figure~\ref{fig:complex3coloring1round} to emphasis the ``circular'' structure of the three components. 

Following the same reasoning as for 0-round algorithms, a 1-round algorithm~\alg\/ induces a name-preserving simplicial map $\delta:\m{P}_{2,\varnothing}^{(1)}\to\LMIS$, as in the second to left diagram in Figure~\ref{fig:ExampleRing}. 

Let us show that such a mapping cannot exist. Since the mapping is name-independent, we consider similarly the mapping of a pair $(p_i,xyz)$ and the mapping of a process view $xyz$. 
For every ordered triplet $(x,y,z)$ of distinct values, $\m{P}_{2,\varnothing}^{(1)}$ contains the following three triangles:
\[
\begin{array}{l}
\{(p_{-1},xyz),(p_0,yzx),(p_1,zxy)\}, \\
 \{(p_{-1},yzx),(p_0,zxy),(p_1,xyz)\}, \;\mbox{and} \\
\{(p_{-1},zxy),(p_0,xyz),(p_1,yzx)\}.
\end{array}
\]
Hence, for each such triplet $(x,y,z)$, one and only one of the three views $xyz, yzx$, and $zxy$ is mapped to~1, while the other two are mapped to~0. Let us assume, w.l.o.g., that $123$ is mapped to~1, while 231 and 312 are mapped to~0. The triangle $\{(p_{-1},212),(p_0,123),(p_1,232)\}$ enforces $212$ and $232$ to be mapped to~0. The triangle $\{(p_{-1},232),(p_0, 321),(p_1,212)\}$ then enforces $321$ to be mapped to~1, and thus 213 and 132 are mapped to~0.  

Now, for every pair $(x,y)$ with $1\leq x<y\leq 3$, there are two triangles 
\[
\{(p_{-1},xyx),(p_0,yxy),(p_1,xyx)\}, \;\mbox{and}\; \{(p_{-1},yxy),(p_0,xyx),(p_1,xyx)\}.
\]
This implies that, for each such pair $(x,y)$, one and only one of the two views $xyx$ and $yxy$ is mapped to~1, while the other is mapped to~0. Thus, in particular, only one of the two views 313 and 131 is mapped to~1, while the other is mapped to~0. It follows that one of the two triangles 
\[
\{(p_{-1},231),(p_0,313),(p_1,132)\}, \;\mbox{and}\; \{(p_{-1},213),(p_0,131),(p_1,312)\}
\]
is mapped to $\{(p_{-1},0),(p_0,0),(p_1,0)\}$, which is not a simplex of  $\LMIS$.

\subparagraph*{Remark.} 

If the input 3-coloring of the ring would be such that the sequence 12321 does not appear as the input colors of five consecutive nodes of~$C_n$, then there would exist a mapping from $\m{P}_{2,\varnothing}^{(1)}$ to $\LMIS$, which in turn demonstrates the existence of a 1-round algorithm under this assumption. More generally, if the sequence $xyzyx$ is guaranteed not to exist in the input 3-coloring for any distinct colors $x,y$, and $z$, then $\delta:\m{P}_{2,\varnothing}^{(1)}\to\LMIS$ defined as
\begin{equation}\label{eq:mapdeltacoloring1round}
\delta(p_i,abc)=\left\{
\begin{array}{ll}
(p_i,1) & \mbox{if $b=x$, or $abc=zyz$; }\\
(p_i,0) & \mbox{if $b=z$, or  $b=y$ with $ac\neq zz$ }
\end{array}\right.
\end{equation}
is a simplicial map. This map induces the 1-round algorithm $\alg$ defined by 
\[
\alg(abc)=\left\{
\begin{array}{ll}
1 & \mbox{if $b=x$, or $abc=zyz$; }\\
0 & \mbox{otherwise. }
\end{array}\right.
\]
That is, nodes colored~$x$ systematically output~1, nodes colored~$z$ systematically output~0, and nodes colored~$y$ output~0 unless they are adjacent to two nodes colored~$z$, in which case they output~1. In fact, only nodes colored~$y$ need to perform a round, the other nodes can decide right away, in zero rounds, based solely on their colors. 

\subparagraph*{Remark.} 

We showed the impossibility of reducing 3-coloring to MIS in a unique round using the impossibility of mapping the complex $\m{P}_{2,\varnothing}^{(1)}$ to the complex $\LMIS$. If one considers merely the \emph{graphs} induced by these two complexes, i.e., their so-called 1-dimensional \emph{skeletons}, then mapping the 1-dimensional  skeleton of  $\m{P}_{2,\varnothing}^{(1)}$  to the  1-dimensional  skeleton of $\LMIS$ is possible by the mapping~$\delta$ of Eq.~\eqref{eq:mapdeltacoloring1round} even if the sequence $xyzyx$ may appear. Indeed, this mapping preserves edges. In particular,  no edges $\{(p_{-1},abc),(p_0,bcd)\}$  (resp., $\{(p_0,abc),(p_1,bcd)\}$) of $\m{P}_{2,\varnothing}^{(1)}$ are mapped by~$\delta$ to the non-edge $\{(p_{-1},1),(p_0,1)\}$ (resp., the non-edge $\{(p_0,1),(p_1,1)\}$) of $\LMIS$. This is to say that, as far as mappings are concerned, the impossibility follows from a contradiction that appears in dimension~2 (i.e., when considering triangles), but not in dimension~1 (i.e., when considering only edges). 

	\subsubsection{The 2-Round Algorithm}
\begin{figure}[htb]
\centering
\includegraphics[scale=0.28]{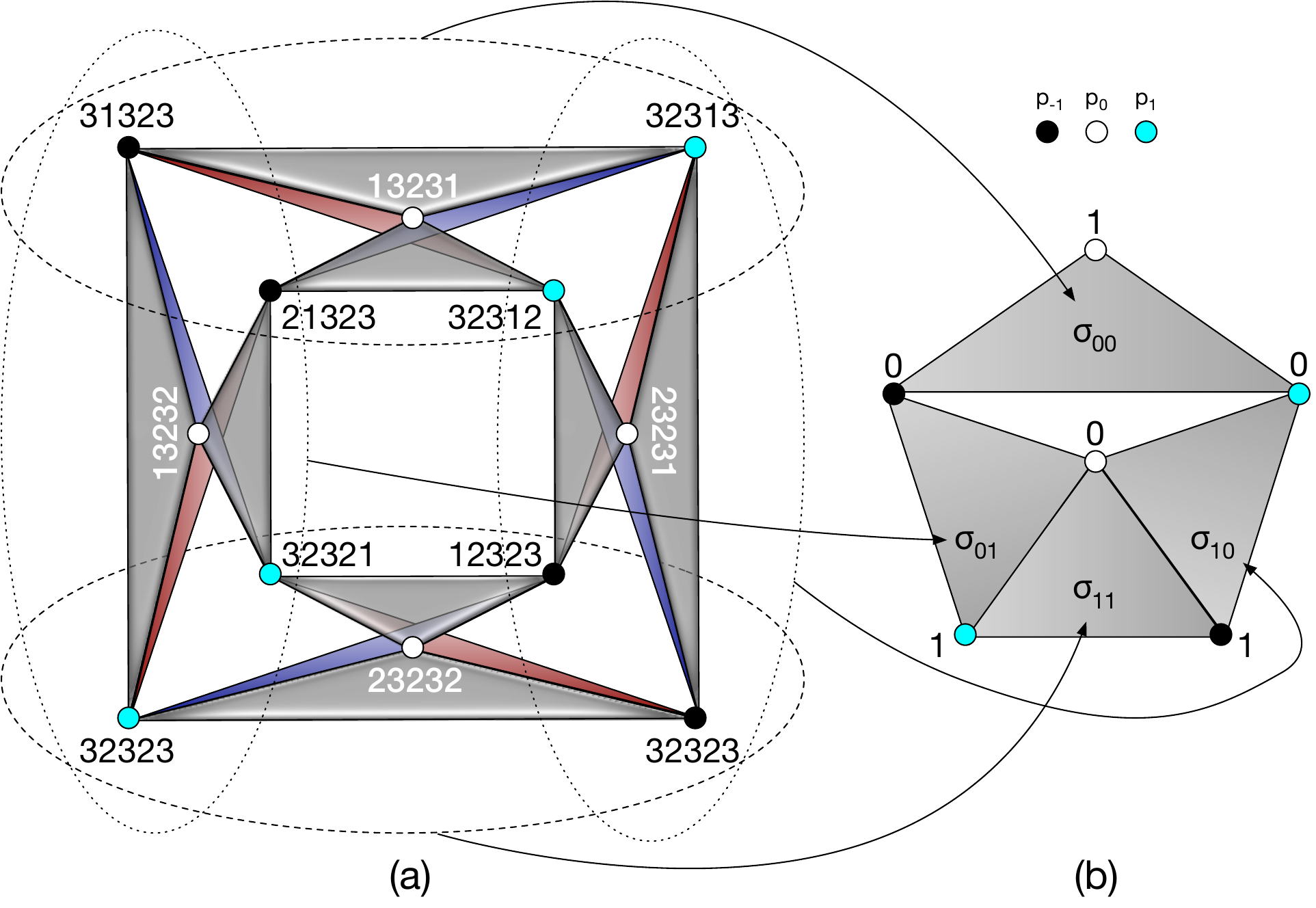}
\caption{\sl (a) The sub-complex $\m{K}_{323}$ of the local protocol complex $\m{P}_{2,\varnothing}^{(2)}$. (b)~The facets of $\LMIS$. } 
\label{fig:complex3coloring2rounds}
\end{figure}

The local protocol complex $\m{P}_{2,\varnothing}^{(2)}$ includes the views of the three nodes $p_{-1}, p_0$, and $p_1$ after two rounds. The vertices of  $\m{P}_{2,\varnothing}^{(2)}$ are of the form $(p_i,c_1c_2c_3c_4c_5)$ with $i\in \{-1,0,1\}$, $c_j\in\{1,2,3\}$ for $1\leq j\leq 5$, and $c_j\neq c_{j+1}$ for $1\leq j<5$. Figure~\ref{fig:complex3coloring2rounds}(a) displays one of the connected components of $\m{P}_{2,\varnothing}^{(2)}$, denoted $\m{K}_{323}$, which includes the four vertices $(p_0,c_1 323 c_5)$, $c_1,c_5\in\{1,2\}$. 
There are 12 disjoint isomorphic copies of this connected component in $\m{P}_{2,\varnothing}^{(2)}$, one for each triplet $c_2,c_3,c_4\in\{1,2,3\}$, $c_2\neq c_3$, and $c_3\neq c_4$. 

Interestingly, each connected component of $\m{P}_{2,\varnothing}^{(2)}$ is isomorphic to each connected component of $\m{P}_{2,\varnothing}^{(1)}$, while there are more connected components in  $\m{P}_{2,\varnothing}^{(2)}$ than in  $\m{P}_{2,\varnothing}^{(1)}$. However, the larger views of the processes provide more flexibility for the mapping from $\m{P}_{2,\varnothing}^{(2)}$ to $\LMIS$ than for the mapping from $\m{P}_{2,\varnothing}^{(1)}$ to $\LMIS$. And indeed, the 2-round anonymous algorithm presented at the end of Section~\ref{subsec:reduction3colMIS} does induce a  name-preserving simplicial map  $\delta:\m{P}_{2,\varnothing}^{(2)}\to\LMIS$. Specifically, the four sub-complexes $\m{K}_{x1y}$, as well as the simplex $\m{K}_{232}$ are entirely mapped to the simplex $\sigma_{00}$ (see Figure~\ref{fig:complex3coloring2rounds}(b) for the labeling of the four facets of $\LMIS$). The two sub-complexes $\m{K}_{1x1}$ are entirely mapped to the simplex  $\sigma_{11}$. The two sub-complexes $\m{K}_{321}$ and $\m{K}_{231}$ are entirely mapped to the sub-complex $\sigma_{01} \cup \sigma_{11}$, and the two sub-complexes $\m{K}_{123}$ and $\m{K}_{132}$ are entirely mapped to the sub-complex $\sigma_{10} \cup \sigma_{11}$. The mapping of the remaining sub-complex $\m{K}_{323}$ is more sophisticated, and illustrates that the simple algorithm showing reduction from 3-coloring to MIS in~\cite{Linial92} is actually topologically non-trivial. 
Indeed, $\m{K}_{323}$ is mapped by the algorithm so that it wraps around the hole in~$\LMIS$. 
This wraparound phenomenon is visualized in Figure~\ref{fig:complex3coloring2rounds}. 

\subsection{General Case with IDs}
So far, we have considered only name-independent algorithms --- algorithms where the nodes do not have IDs or do not use them. 
Recall that the name~$i\in\{-1,0,1\}$ of a process $p_i$ is external to the system, and is used only for analyzing the ability to solve tasks. 
The presence of IDs given to the nodes adds power to the distributed algorithms, as the output of a process is not only a function of the observed colors in its neighborhood, but also of the observed IDs. In particular, after one round, a process $p$ is not only aware of a triplet of colors $(c_1c_2c_3)$, but also of a triplet of distinct IDs $(x_1x_2x_3)$. 

	\subsubsection{Impossibility in Zero Rounds  with IDs}
\begin{figure}[htb]
\centering
\includegraphics[scale=0.4]{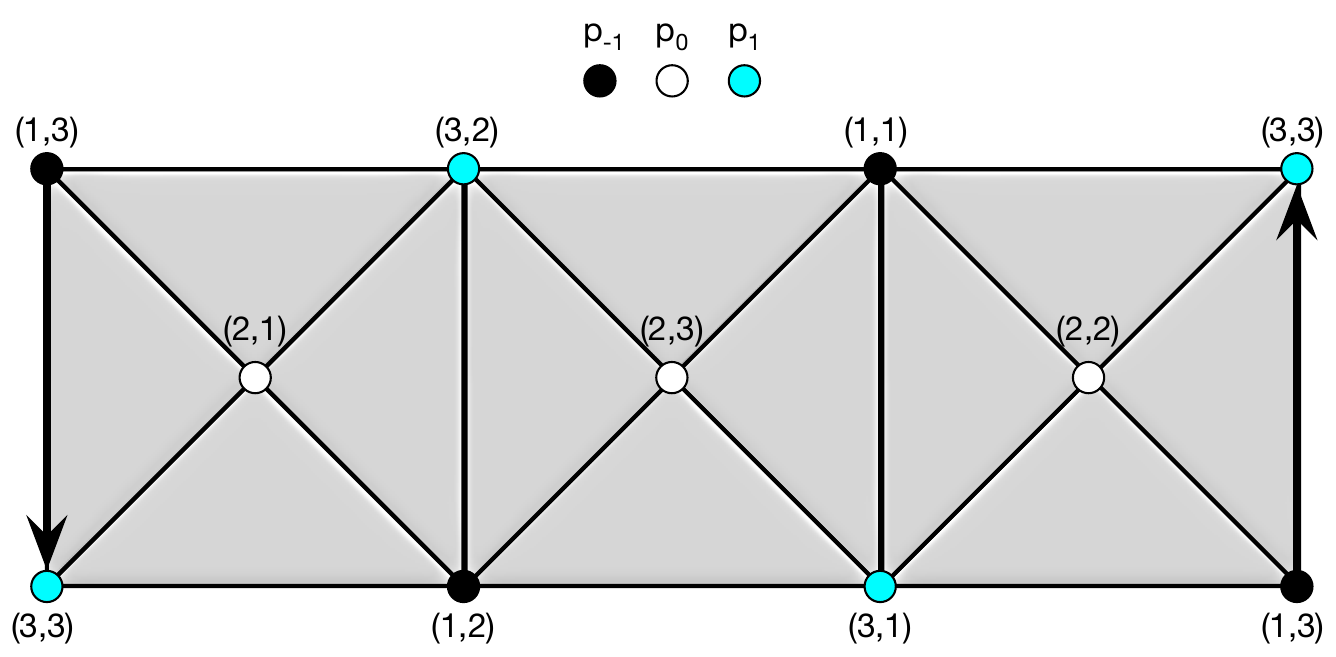}
\caption{\sl Local input complex $\LCOL_{2,\text{\rm fix}}$ of 3-coloring in the ring with fixed  IDs in $\{1,2,3\}$.}
\label{fig:complex3coloringID}
\end{figure}

A local input complex $\LCOL_{2,\text{fix}}$ for 3-coloring with \emph{fixed} IDs is displayed on Figure~\ref{fig:complex3coloringID}.
Each vertex is a pair $(p_i,(x,c))$, where $p_i$, $i\in\{-1,0,1\}$, is the name of a process, $x\in\{1,2,3\}$~is an ID, and $c\in\{1,2,3\}$ is a color. In this figure, it is assumed that $p_{-1}$ is systematically given ID~1, $p_0$ is systematically given ID~2, and $p_1$ is systematically given ID~3. 
This complex is only a small part of a complex describing a colored ring, where the number of IDs is larger and any process can be given any ID.

\subparagraph{Remark.} The complex $\LCOL_{2,\text{fix}}$ is not the complex $\LCOL_{2,[3]}$ as specified  on Figure~\ref{fig:commdiag}, because $\LCOL_{2,[3]}$ assumes that every process $p_i$, $i\in\{-1,0,1\}$, can take every possible ID in $[3]=\{1,2,3\}$. 
In fact, $\LCOL_{2,\text{fix}}$ can be appropriately mapped to the local complex $\LMIS$. 
A trivial name-preserving name-independent mapping is, for every $i\in\{-1,0,1\}$, 
\begin{equation}
\delta(p_i,(x,c))=\left\{
\begin{array}{ll}
(p_i,1) & \mbox{if $x=2$}\\
(p_i,0) & \mbox{otherwise.}
\end{array}\right.
\label{eq:trivialmapping}
\end{equation}

We stress that this does not imply the existence of an algorithm reducing 3-coloring to MIS, as in reality the IDs are not fixed.
To show impossibility of reducing 3-coloring to MIS in zero rounds, a more sophisticated complex must be considered, in which \emph{IDs are not fixed a priori}. 

\begin{figure}[htb]
\centering
\includegraphics[scale=0.38]{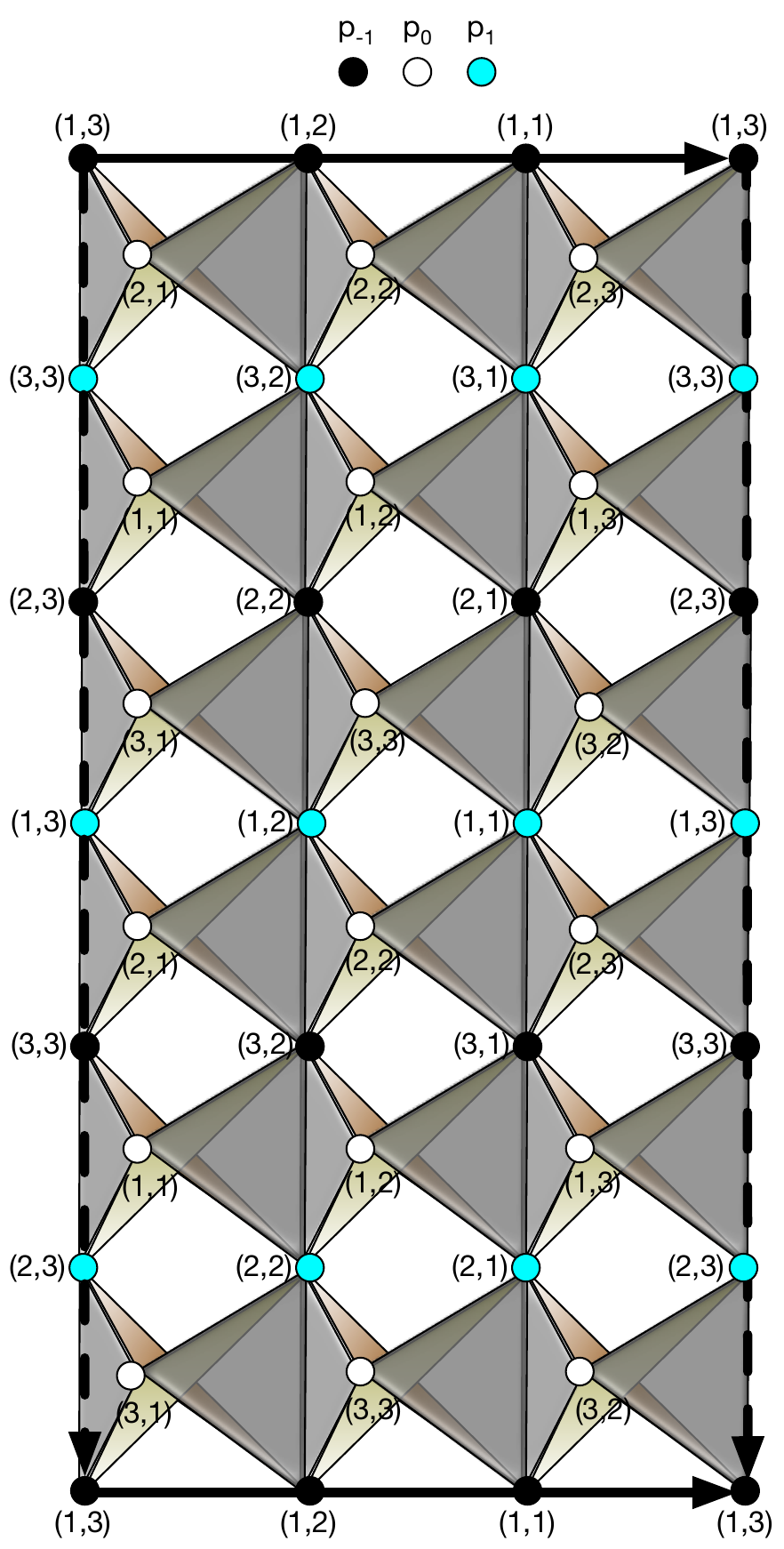}
\caption{\sl Local input complex $\LCOL_{2,[3]}$ of 3-coloring in the ring with arbitrary IDs  in $\{1,2,3\}$.}
\label{fig:complex3coloringIDall}
\end{figure}

First, let us consider the case where $p_{-1}$, $p_0$, and $p_1$ take \emph{any} assignment of unique IDs in $\{1,2,3\}$, and not posses fixed IDs as above. 
The resulting input complex $\LCOL_{2,[3]}$ is displayed on Figure~\ref{fig:complex3coloringIDall}. The vertices are arranged on a grid, and the figure wraps around in a way similar to a torus. The four triangles forming cones centered at vertices $(p_0,(x,c))$ with $(x,c)\in\{1,2,3\}^2$ are ``twisted'', and each of these latter vertices is appearing twice in the figure, for allowing the figure to be displayed as a torus. 
(The specific ID assignment that appeared in Figure~\ref{fig:complex3coloringID} is the upmost  part of Figure~\ref{fig:complex3coloringIDall}, twisted.)
Despite its apparent complexity, the complex $\LCOL_{2,[3]}$ can  be appropriately mapped to $\LMIS$, using again the simplicial map of Eq.~\eqref{eq:trivialmapping}.  This shows that \emph{more IDs must be considered to show impossibility}.

Since the simplicial maps $\delta$ induced by the potential algorithms are name-preserving, they actually act on pairs $(x,c)$ where~$x$ is an ID and~$c$ is a color, i.e., $\delta(p,(x,c))=(p,\hat{\delta}(x,c))$ for some  $\hat{\delta}$. For brevity, we identify $\hat{\delta}$ with~$\delta$. Let us assume that the IDs are from $\{1,\dots,R\}$, for some $R\geq 4$. That is, we consider now $\LCOL_{2,[R]}$ for $R\geq 4$. By the pigeon-hole principle, there exists a set $I_1\subseteq \{1,\dots,R\}$ with $|I_1|\geq R/2$ such that, for every $x,x'\in I_1$, $\delta(x,1)=\delta(x',1)$. Therefore, again by  the pigeon-hole principle, there exists a set $I_2\subseteq I_1$ with $|I_2|\geq |I_1|/2$ such that, for every $x,x'\in I_2$, $\delta(x,2)=\delta(x',2)$. Finally, there exists a set $I_3\subseteq I_2$ with $|I_3|\geq |I_2|/2$ such that, for every $x,x'\in I_3$, $\delta(x,3)=\delta(x',3)$. Therefore,  there exists a set $I\subseteq \{1,\dots,R\}$ with $|I|\geq R/8$ such that, for every $x,x'\in I$, $\delta(x,1)=\delta(x',1)$, $\delta(x,2)=\delta(x',2)$, and $\delta(x,3)=\delta(x',3)$. 
Whenever $R\geq 24$, the set $I$ has size at least~$3$. Consider the sub-complex  $\LCOL_{2,[R]}'$ of $\LCOL_{2,[R]}$  induced by the three smallest IDs in~$I$ --- this sub-complex is isomorphic to $\LCOL_{2,\varnothing}$ (Figure~\ref{fig:complex3coloring}). More importantly, the mapping from $\LCOL_{2,[R]}'$ to $\LMIS$ depends only on the colors and not on the IDs, by the choice of~$I$. Hence, if there was a mapping from $\LCOL_{2,[R]}'$ to $\LMIS$, then there would exist a mapping from $\LCOL_{2,\varnothing}$ to $\LMIS$, which we know does not exist. 

It follows that there are no mappings from $\LCOL_{2,[24]}=\m{P}^{(0)}_{2,[24]}$ to $\LMIS$ --- see the second to right diagram in Figure~\ref{fig:ExampleRing}. In other words, if the IDs are picked from a set of at least 24 values, then 3-coloring cannot be reduced to MIS in zero rounds. 

\subparagraph*{Remark.}
We have presented the pigeon-hole argument in detail because it can be generalized and give a good intuition for the general case. However, the impossibility of reducing 3-coloring to MIS in zero rounds can actually be established  by letting nodes taking IDs in a much smaller set, namely in the set $\{1,2,3,4\}$. Indeed, the resulting complex $\LCOL_{2,[4]}=\m{P}^{(0)}_{2,[4]}$ cannot be mapped to $\LMIS$ by a name-preserving name-independent simplicial map. To see why, let us assume for contradiction that such a mapping exists. 
For every ID~$x$, $\LCOL_{2,[4]}$ includes triangles in which no vertex has ID~$x$. 
Similarly, for every color~$c$, $\LCOL_{2,[4]}$ includes triangles in which no vertices have color~$c$. 
It follows that the pre-image of $(p_0,1)$ must include at least two vertices $(p_0,(x,c_x))$ and  $(p_0,(x',c_{x'}))$ with $x\neq x'$ for some (possibly identical) colors $c_x$ and~$c_{x'}$, and at least two vertices $(p_0,(x_c,c))$ and  $(p_0,(x_{c'},c'))$ with $c\neq c'$ for some  (possibly identical) IDs $x_c$ and~$x_{c'}$. 
As a consequence, there are two distinct IDs $x$ and $x'$, and two distinct colors $c$ and $c'$ such that  $(p_0,(x,c))$ and  $(p_0,(x',c'))$ are both in the pre-image of $(p_0,1)$. This yields a contradiction as the simplex $\{(p_0,(x,c)),(p_1,(x',c'))\}$ would then be mapped to $\{(p_0,1),(p_1,1)\}$, which is not a simplex in~$\LMIS$.

	\subsubsection{Impossibility in One Rounds with IDs}
We reduce the case with IDs to the case without IDs in a way similar to the case of zero rounds, by using Ramsey's theorem instead of the basic pigeon-hole principle, following the lines of~\cite{NaorS95}. Recall that Ramsey's theorem states the following. Given a set~$X$ and a non-negative integer~$s$, let  $X\choose s$ denote the set of all subsets of $X$ with exactly $s$ elements. In particular, $X\choose s$ has cardinality~$|X|\choose s$. 

\begin{theorem}[Ramsey's Theorem~\cite{GRSS2015}]\label{theo:Ramsey}
	For all $r,s,t \in \mathbb{N}$, there exists $R=R(r,s,t)$ such that, for every set $X$, and for every partition of $X\choose s$ into $r$ classes, if $|X|\geq R$, then one of the classes contains all elements of  $Y\choose s$, for some set $Y\subseteq X$ with $|Y|\geq t$. \end{theorem}

We consider the 1-round protocol complex with IDs in a finite set $X$ and at least 5~elements, denoted by $\m{P}_X$. That is, $\m{P}_X=\m{P}^{(1)}_{2,[k]}$ with $k=|X|$. The vertices of this complex are of the form $(p_i,(xyz,abc))$ where $i\in\{-1,0,1\}$, $\{x,y,z\}\in {X\choose 3}$, and $a,b,c\in\{1,2,3\}$ with $a\neq b$ and $b\neq c$. The facets of $\m{P}_X$ are of the form
\[
F_{x'xyzz',a'abcc'}=\{(p_{-1},(x'xy,a'ab)),(p_0,(xyz,abc)),(p_1,(yzz',bcc'))\},
\]
with $\{x',x,y,z,z'\}\in {X\choose 5}$, and $a',a,b,c,c'\in\{1,2,3\}$ with $a'\neq a\neq b\neq c \neq c'$. 

A name-preserving name-independent simplicial map $\delta:\m{P}_X\to \LMIS$ induces a labeling of the pairs $(xyz,abc)$ with labels in $\{0,1\}$, where $xyz$ is an ordered triplet of distinct IDs, and $abc$ is an ordered triplet of colors in $\{1,2,3\}$ with $a\neq b$ and $b\neq c$. It follows that $\delta$ induces a labeling of the ordered triplets $xyz$  of distinct IDs by labels in $\{0,1\}^{12}$, by applying $\delta$ to the 12 possible choices of color triplets. More specifically, let us lexicographically order the $12$ different ordered triplets of colors, and let us denote by $C_1,\dots,C_{12}$ this lexicographic  ordering. We aim at labeling sets of IDs, not ordered triplets of IDs. Let  $S=\{x,y,z\}$  be a set of distinct IDs, and assume that $x<y<z$. The set $S$ is assigned the label equal to the binary vector in $\{0,1\}^{12}$ whose $i$-th entry is equal to $\delta(p_0,(xyz,C_i))$.  

Let $r=2^{12}$, $s=3$, and $t=5$. By Ramsey's Theorem, by taking the IDs in the set $X=\{1,\dots,R\}$ with $R=R(r,s,t)$, there exists a set $Y$ of $t=5$ IDs such that, for every two sets $\{x,y,z\}$ and $\{x,'y',z'\}$ of IDs in~$Y$, with $x<y<z$ and $x'<y'<z'$, and for every ordered triplet $abc$ of colors, we have
\[
\delta(p_0,(xyz,abc))=\delta(p_0,(x'y'z',abc)). 
\]
More generally, by name-independence, for such IDs and colors, we actually have 
\[
\delta(p_i,(xyz,abc))=\delta(p_j,(x'y'z',abc))
\]
for every $i,j\in\{-1,0,1\}$. Let $\m{P}_Y$ be the sub-complex of the 1-round protocol complex $\m{P}_X$ induced by the vertices with IDs in~$Y$ ordered in increasing order. That is, we keep in $\m{P}_Y$ solely the vertices of $\m{P}_X$ of the form $(p_i,(xyz,abc))$ with $\{x,y,z\}\subset Y$ and $x<y<z$. By construction of $Y$, $\delta$ is name-independent on $\m{P}_Y$. 

Now, recall the protocol complex $\m{P}^{(1)}_{2,\varnothing}$ displayed on Figure~\ref{fig:complex3coloring1round}. Let us define the map $\delta':\m{P}^{(1)}_{2,\varnothing}\to \LMIS$ by $\delta'(p_i,abc)=\delta(p_i,(xyz,abc))$ where $\{x,y,z\}\subset Y$ and $x<y<z$. Note that $\delta'$ is well defined as $\delta$ is name-independent on $Y$. Moreover, assuming $\delta:\m{P}_X\to\LMIS$ is simplicial yields that $\delta':\m{P}^{(1)}_{2,\varnothing}\to \LMIS$ is simplicial as well. We have seen in Section~\ref{subsubsec:no1roundwithoutID} that such a simplicial mapping does not exist. 

\label{subsubsec:impossibility-1-round-with-ID}

\subsection{Wrap Up}

This section provided an illustration of the fact that the complexity of LCL tasks can be analyzed by considering finite simplicial complexes, even if the tasks were defined for arbitrarily large networks, whose nodes are assigned IDs from an arbitrarily large range of values. The next section provides a formalization of the examples in this section, and generalize them to establish our main result. 
\section{Topology of LCL Tasks}
We now show how to study a general LCL task in the \local\/ model by representing it in topological terms. 
For this,
we define the input and output complexes, the relation between them, and the protocol complexes for LCL tasks in  the \local\/ model. 
Let $S_d$ be the star of $d+1$ nodes, whose center node is named $p_0$, and the leaves are named $p_i$, for $i=1,\dots,d$. We consider algorithms for classes $\m{G}\subseteq\m{G}_d$ of graphs, where $\m{G}_d$ denotes the set of all $d$-regular connected simple graphs.  

\begin{definition}\label{def:IandOforLCL}
Let $T=(\m{L}_{in},\m{S}_{in},\m{L}_{out},\m{S}_{out})$ be an LCL task for $\m{G}\subseteq \m{G}_d$. The input complex~$\m{I}_d$  (resp., output complex~$\m{O}_d$) associated with $T$ is the complex where 
$
\{(p_i,x_i) : i\in \{0,\dots,d\}\}
$
is a facet of~$\m{I}_d$ (resp., a facet of~$\m{O}_d$) if $x_i\in \m{L}_{in}$ (resp., $\m{L}_{out}$) for every~$i\in\{0,\dots,d\}$, and the labeled star resulting from assigning label $x_i$ to the node $p_i$ of $S_{d}$ for every $i\in \{0,\dots,d\}$ is in $\m{S}_{in}$ (resp.,~$\m{S}_{out}$). 
The carrier map $
\Delta:\m{I}_d\to 2^{\m{O}_d}$ is defined simply by
$\Delta(F)=\{\mbox{all facets of}\;\m{O}_d\}$ for every facet $F$ of $\m{I}_d$.
\end{definition}
Note that with this definition at hand, we can write the same LCL task $T$ both as $T=(\m{L}_{in},\m{S}_{in},\m{L}_{out},\m{S}_{out})$ and as $T=(\m{I}_d,\m{O}_d,\Delta)$.
The interpretation will be clear from the context.

If the considered LCL task $T$  imposes constraints on the correctness of the outputs as a function of the inputs, as in list-coloring, then the carrier map 
$
\Delta:\m{I}_d\to 2^{\m{O}_d}
$
is not as above, and instead it specifies for each facet $F\in\m{I}_d$ the facets $\Delta(F)$ which are legal with respect to~$F$. For instance, in the case of list-coloring where each $x_i$ is a list of colors, for every facets $F=\{(p_i,x_i) : i\in \{0,\dots,d\}\}\in \m{I}$ and $F'=\{(p_i,y_i) : i\in \{0,\dots,d\}\}\in \m{O}_d$ we have 
\[
F' \in \Delta(F) \; \iff \; \forall  i\in \{0,\dots,d\}, \; y_i\in x_i. 
\]
Note that we do not have to require $y_i\neq y_0$ here, since global states where $y_i= y_0$ are not simplices of $\m{O}_d$, and so no simplex can be mapped to them. 

\subparagraph*{Mutually compatible views.}

Let $t\geq 0$, and let us fix a graph $G=(V,E)$ in $\m{G}\subseteq\m{G}_d$. 
In $t$ rounds, every node in~$G$ acquires a \emph{view}~$w$, whose structure is isomorphic to a radius-$t$ ball in~$G$ centered at that node, including the input labels and the IDs of the nodes in the ball. 
The number of nodes in a view after $t$ rounds is at most $N(d,t)$ where, for every $t\geq 0$, $N(d,t)=1+d+d(d-1)+\dots+d(d-1)^{t-1}$, that is, 
\[
N(d,t)=
\begin{cases}
	1+2t & \text{if } d=2, \\
	1+d\,\frac{(d-1)^t-1}{d-2} & \text{otherwise.}
\end{cases}
\]
This number of nodes is exactly $N(d,t)$ if all graphs in $\m{G}$ have girth at least $2t+1$,
(i.e., if the graphs have no cycles of less than $2t+1$ nodes), and every $t$-round view is a $d$-regular tree.
An ordered collection $w_0,\dots,w_d$ of views at distance~$t$ forms a collection of \emph{mutually compatible} views for $\m{G}$ if there exists a graph $G\in \m{G}$, an assignment of input labels and IDs to the nodes of $G$, and a star $S$ in $G$ with center~$v_0$ and leaves $v_1,\dots,v_d$, such that $w_i$ is the view of $v_i$ in $G$ after $t$ rounds, for $i=0,\dots,d$. 

\begin{definition}\label{def:protocol-complex-with-IDs}
Let $T$ be an LCL task for $\m{G}\subseteq \m{G}_d$, and let $t\geq 0$. The $t$-round protocol complex associated with $T$ for a finite set $X$ of IDs, is the complex~$\m{P}^{(t)}_{d,X}$ where 
$
F=\{(p_i,w_i) : i\in \{0,\dots,d\}\}
$
is a facet of~$\m{P}^{(t)}_{d,X}$ if $w_0,\dots,w_d$ is an ordered collection of mutually compatible views at distance $t$ for $\m{G}$.
\end{definition}

The special case $t=0$ corresponds to $\m{P}^{(0)}_{d,X}=\m{I}_{d,X}$ where $\m{I}_{d,X}$ in the input complex $\m{I}_d$ extended with IDs in~$X$. In this specific case, mutual compatibility requires the additional condition that the processes $p_0,\dots,p_d$ are given distinct IDs in~$X$. Two mappings from $\m{I}_{d,X}$ play a crucial role. The first is the simplicial map
\[
\pi:\m{I}_{d,X}\to \m{I}_d
\]
defined by $\pi(p_i,(\id,x))=(p_i,x)$ for every $i=0,\dots,d$, every $\id\in X$, and every $x\in\m{L}_{in}$. The second is the carrier map 
\[
\Xi_t:\m{I}_{d,X}\to 2^{\m{P}_{d,X}^{(t)}}
\]
that specifies, for each facet $F\in\m{I}_{d,X}$, the set $\Xi_t(F)$ of facets which may result from~$F$ after $t$~rounds of computation in graphs in~$\m{G}$. Specifically, they are merely the facets of $\m{P}_{d,X}^{(t)}$ for which the views $w_0,\dots,w_d$ are compatible with the IDs of $p_0,\dots,p_d$ in~$F$. 
While formally $\Xi_t$ is defined on all simplices, note that defining $\Xi_t$ on facets is sufficient, as it can easily be extended to all other simplices.

Our main result is an analog of the generic lemma (see Lemma~\ref{lem:generic}), but involving local complexes, even for tasks defined on arbitrarily large networks, and for arbitrarily large sets of IDs. Specifically, in the statement below, the range~$[R]=\{1,\dots,R\}$ of IDs depends only on the number of rounds~$t$ of the algorithm, the maximum degree~$d$ of the network, and the respective sizes  $|\m{L}_{in}|$ and $|\m{L}_{out}|$ of the input and output labels. That is, the range $[R]$ is independent of the size of the network, as well as of the range of IDs. Theorem~\ref{theo:main} is the formal version our main result sketched in Theorem~\ref{theo:main-simplified}. 

\begin{theorem}\label{theo:main}
Let $T=(\m{I}_d,\m{O}_d,\Delta)$ be an LCL task for $\m{G}\subseteq \m{G}_d$, and let $t\geq 0$. 
\begin{itemize}
\item If there exists a distributed algorithm solving $T$ in $t$ rounds in the  \local\/ model then, for every $R\geq N(d,t+1)$, there is a name-preserving name-independent simplicial map $\delta:\m{P}_{d,[R]}^{(t)}\to \m{O}_d$ such that, for every facet $F\in\m{I}_{d,[R]}$,  $\delta(\Xi_t(F))\subseteq \Delta(\pi(F))$. 
\item There exists $R\geq N(d,t+1)$ satisfying that, if there is a name-preserving name-independent simplicial map $\delta:\m{P}_{d,[R]}^{(t)}\to \m{O}_d$ such that, for every facet $F\in\m{I}_{d,[R]}$,  $\delta(\Xi_t(F))\subseteq \Delta(\pi(F))$, then there is a distributed algorithm solving $T$ in $t$ rounds in the  \local\/ model. 
\end{itemize}
\end{theorem}

\proof
Let us fix an LCL task $T=(\m{L}_{in},\m{S}_{in},\m{L}_{out},\m{S}_{out})=(\m{I}_d,\m{O}_d,\Delta)$ for $\m{G}$, and $t\geq 0$. Let \alg\/ be a $t$-round algorithm for~$T$. For any finite set $X$ of IDs with $|X|\geq N(d,t+1)$, let us define $\delta_X:\m{P}_{d,X}^{(t)}\to \m{O}_d$ by 
\[
\delta_X(p_i,w_i)=(p_i,\alg(w_i)),
\]
for every $i=0,\dots,d$. By construction, $\delta_X$ is name-preserving and name-independent. To show that $\delta_X$ is simplicial, let 
\[
F'=\{(p_i,w_i) : i\in \{0,\dots,d\}\}
\]
be a facet of $\m{P}_{d,X}^{(t)}$. This facet is mapped to 
\[
\delta_X(F')=\{(p_i,\alg(w_i)) : i\in \{0,\dots,d\}\}.
\] 
Since $\alg$ solves $T$, every output $\alg(w_i)$ is in $\m{L}_{out}$, and the labeled star resulting from assigning label $\alg(w_i)$ to the node $p_i$ of the star $S_{d}$, for every $i\in \{0,\dots,d\}$, is in~$\m{S}_{out}$. It follows that $\delta_X(F')$ is a facet of $\m{O}_d$, and thus $\delta_X$ is simplicial. Moreover, if the facet $F'$ belongs to the image $\Xi_t(F)$ of a facet $F$ of  $\m{I}_{d,X}$, since $\alg$ solves~$T$, it follows that $\delta_X(F')\in \Delta(\pi(F))$ as desired. 

So, the existence of an algorithm $\alg$ guarantees the existence of a simplicial map $\delta_X$ satisfying the requirements of the theorem for \emph{every} large enough set $X$ of IDs. We now show that, to guarantee the existence of an algorithm, it is sufficient to guarantee the existence of a simplicial map $\delta_X$ just for \emph{one} specific set~$X=[R]$. 

In order to identify $R$, we follow the same lines as in the impossibility proof in Section~\ref{subsubsec:impossibility-1-round-with-ID}, using Ramsey's theorem (cf. Theorem~\ref{theo:Ramsey}).
Note that the number of possible balls of radius~$t$ in graphs of~$\m{G}$ is finite, and depends only on~$t$ and~$d$.  Given such a ball~$B$, there are finitely many ways of assigning input labels to the vertices of $B$. The number of assignments depends only on the structure of $B$, and on $|\m{L}_{in}|$. (It may also depend on~$\m{S}_{in}$, but in the worst case, all assignments are possible.) Let us enumerate all the labeled balls in~$\m{G}$ as 
\[
B^{(1)},\dots,B^{(k)}. 
\]
The number $k$ of such labeled balls depends only on $d$, $t$, and $|\m{L}_{in}|$. (It may also depend on $\m{G}$, but it is upper bounded by a function of $d$, $t$, and $|\m{L}_{in}|$.) 

For every labeled ball $B^{(i)}$, $i=1,\dots,k$, let $\nu_i=|B^{(i)}|$. Let us rank the vertices of $B^{(i)}$ arbitrarily from~$1$ to $\nu_i$, and let $\Sigma_i$ be the set of all permutations of $\{1,\dots,\nu_i\}$. To every $\pi\in\Sigma_i$ corresponds a labeled ball $B^{(i)}_\pi$ in which the rank of the vertices is determined by~$\pi$. 

Now, let $X$ be a finite set of IDs with $|X|\geq N(d,t+1)$.
We lower bound $|X|$ by $N(d,t+1)$ and not $N(d,t)$ because we want to consider the behavior of a simplex, i.e., balls of radius $t$ around a process $p_0$ and around each of its neighbors $p_1,\dots,p_d$.
We consider all possible identity-assignments with IDs in~$X$ to the nodes of the labeled balls  with ranked vertices, $B^{(i)}_\pi$, $i=1,\dots,k$, $\pi\in\Sigma_i$, as follows. 

For every $S\subseteq X$ with $|S|=N(d,t)$, let us order the IDs in $S$ in increasing order. Given a ranked labeled ball $B^{(i)}_\pi$, i.e., a labeled ball $B^{(i)}$ whose vertices are ranked by some permutation~$\pi\in\Sigma_i$, the IDs in $S$ are assigned to the nodes of $B^{(i)}_\pi$ by assigning the $j$th smallest ID in $S$ to the node ranked $\pi(j)$ in $B^{(i)}_\pi$, for $j=1,\dots,\nu_i$. 

By picking all $i=1,\dots,k$, all $\pi\in\Sigma_i$, and all $S\subseteq X$, we obtain all possible views resulting from performing a $t$-round algorithm in $\m{G}$ with IDs taken from $X$. Let us order these views as 
\[
w^{(1)},\dots,w^{(h)},
\] 
where the views induced by $B^{(1)}$ are listed first, then the views induced by $B^{(2)}$, etc., until the views induced by $B^{(k)}$. Moreover, for a given  $i\in\{1,\dots,k\}$, the views corresponding to the labeled ball~$B^{(i)}$ are listed according to the lexicographic order of the permutations in~$\Sigma_i$. Note that the number $h$ of views depends only on $d$, $t$, $|\m{L}_{in}|$, and $|X|$. 

Each set $S$ is then ``colored'' by  
\[
c(S)=(\delta_X(p_0,w^{(1)}),\dots,\delta_X(p_0,w^{(h)})) \in \{1,\dots,|\m{L}_{out}|\}^h.
\]
In this way, the set $X\choose N(d,t)$ is partitioned into $|\m{L}_{out}|^h$ classes. 
Thanks to Ramsey's Theorem (see Theorem~\ref{theo:Ramsey}), by taking set
\[
X=[R] \; \mbox{with} \; R=R(a,b,c) \; \mbox{for} \; 
a=|\m{L}_{out}|^h, \; 
b=N(d,t),\mbox{and} \; 
c=N(d,t+1),
\]
we are guaranteed that there exists a set $Y$ of at least $N(d,t+1)$ IDs such that every two sets $S$ and $S'$ of $N(d,t)$ IDs in~$Y$ are given the same color $c(S)=c(S')$. In other words, for any ball $B$ of radius~$t$ in a graph from $\m{G}$, and for every valid assignment of input values to the nodes of $B$, if one assigns the IDs in $S$ and $S'$ in the same manner (i.e., the $i$th smallest ID of $S$ is assigned to the same node as the $i$th smallest ID of $S'$), then 
\[
\delta_X(p_0,w)=\delta_X(p_0,w'), 
\]
where $w$ and $w'$ are the views resulting from assigning IDs from $S$ and $S'$ to the nodes, respectively. 

Now, let us define the following $t$-round algorithm \alg\/ for $T$;
in fact, this is precisely the order-invariant algorithm constructed in~\cite{NaorS95}. 
To this end, we assume that the set $Y$ is pre-computed and hard-wired to the algorithm.
Every node~$v$ collects the data available in its centered ball $B=B_G(v,t)$ of radius~$t$ in the actual graph $G\in\m{G}$, where $B$ contains both IDs and input values. 
Node~$v$ reassigns the IDs to the nodes of $B$ by using the $|B|$ smallest IDs in $Y$, and assigning these IDs to the nodes of $B$ in the order respecting the order of the actual IDs assigned to the nodes of~$B$. 
Then, node~$v$ considers the view $w$ after reassignment of the IDs, and outputs
\[
\alg(w)=\delta_X(p_0,w).
\]
Note that $\delta_X$ returns values in $\m{L}_{out}$, and thus $\alg$ is well defined. 

To show correctness, let us consider a star $v_0,\dots,v_d$ centered at $v_0$ in some graph $G\in\m{G}$. Performing $\alg$ in $G$, each of these $d+1$ nodes acquires a view of radius~$t$. These views are mutually compatible. Let us reassign the IDs in the ball of radius $t+1$ centered at $v_0$ in $G$, using the at most $N(d,t+1)$ smallest IDs in~$Y$, and assigning these IDs to the nodes of the ball $B$ of radius $t+1$ centered at $v_0$, in the order respecting the order of the actual IDs assigned to the nodes of $B$. The resulting views $w_0,\dots,w_d$ of the $d+1$ nodes $v_0,\dots,v_d$ remain mutually compatible. It follows that if these $d+1$ nodes would output $\delta_X(p_0,w_0),\dots,\delta_X(p_d,w_d)$, respectively, then the resulting star would be good. We claim that this is exactly what occurs with~\alg. 

Indeed, first, $\delta_X$ is name-independent, and thus $\delta_X(p_0,w)=\delta_X(p_i,w)$ for every $i=1,\dots,d$. Second, and more importantly, by the construction of $Y$, the actual \emph{values} of the IDs do not matter, but solely their \emph{relative order}. The reassignment of IDs performed at each of the nodes $v_0,\dots,v_d$ is different from the reassignment of IDs in the ball $B$ of radius~$t+1$ around~$v_0$, but the relative order of these IDs is preserved as it is governed by the relative order of the original IDs in~$B$. As a consequence, the nodes of the star $S_d$ consisting of $p_0$ and its $d$ neighbors correctly output $\delta_X(p_0,w_0),\dots,\delta_X(p_d,w_d)$ in $\alg$, as desired.  
\qed

\section{Application to Coloring the Ring}

In this section, we show a concrete application of Theorem~\ref{theo:main}, by reproving the celebrated result by Linial~\cite{Linial92} regarding 3-coloring the $n$-node ring. 
This results was later re-proven in a simplified way~\cite{LaurinharjuS14}, basically using the original arguments  but providing a purely \emph{combinatorial perspectives} on them. Also, \cite{Balliu0HORS19,Brandt19} recently introduced a general round-reduction \emph{operational} technique for deriving lower bounds in the \local\/ model. 
In this section, we provide a \emph{topological perspective} on lower bounds in the \local\/ model. 
Specifically, we prove the following corollary of Theorem~\ref{theo:main}. 

\begin{corollary}\label{cor:reproving-linial}
Let $t\geq 1$, $k\geq 2$, $n\geq 1$, and $N\geq n$. If there is a $t$-round algorithm for $k$-coloring $C_n=(v_1,\dots,v_n)$ when the IDs in $[N]$ are assigned to consecutive nodes $v_i,v_{i+1}$, $i\in\{1,\dots,n-1\}$, in increasing order of their indices, then there is a $(t-1)$-round algorithm for $2^{2^k}$-coloring $C_n$ under the same constraints on the ID assignment. 
\end{corollary}

\proof
Observe first that the value of $R$ in Theorem~\ref{theo:main} is non-decreasing with~$t$. Therefore, we fix the $R$ defined for $t$, and use the same~$R$ for $t-1$. Also, since we solely focus on the ring in the proof, we fix $d=2$ and omit it from the notation of the relevant complexes. 
By Theorem~\ref{theo:main}, since there is a $t$-round algorithm for $k$-coloring the ring, there is a name-preserving name-independent simplicial map $\delta:\m{P}_{[R]}^{(t)}\to \m{O}_k$ with the property that, for every facet $F\in\m{I}_{[R]}$,  $\delta(\Xi_t(F))\subseteq \Delta(\pi(F))$, where $\Delta$ is the carrier map specifying $k$-coloring and $\m{O}_k$ is the output complex for $k$-coloring. 
Also, $\m{I}_{[R]}$ is the input complex with no inputs to the vertices, apart from their IDs in $[R]$. More precisely, in~$\m{I}_{[R]}$, since the IDs are assigned in increasing order, we restrict our interest to nodes $p_0$ which are neither $v_1$ nor $v_n$ and to facets~$F$ of the form
\[
F=\big \{(p_{-1},x),(p_0,y),(p_1,z)\big\} \; \mbox{with} \; x,y,z \in [R], \; \mbox{and} \; x < y < z. 
\]
The same restriction on the IDs applies to the facets of $\m{P}_{[R]}^{(t)}$. 

\subparagraph{Sketch of the arguments.} 

Our aim is to find $\delta':\m{P}_{[R]}^{(t-1)}\to \m{O}_{2^{2^k}}$ where $\m{O}_{2^{2^k}}$ is output complex for $2^{2^k}$-coloring~$C_n$.  For this purpose, we follow the approach illustrated on Figure~\ref{fig:f**f**}. That is, first, we identify a  functor~$\Phi$ on a category corresponding to a subclass of simplicial complexes. From the simplicial map $\delta:\m{P}_{[R]}^{(t)}\to \m{O}_k$, we derive the simplicial map $\Phi(\delta):\Phi(\m{P}_{[R]}^{(t)}) \to \phi(\m{O}_k)$. Then we show that  $\Phi(\m{O}_k) \subseteq \m{O}_{2^{2^k}}$ as sub-complex, and therefore $\Phi(\delta)$ maps $\Phi(\m{P}_{[R]}^{(t)})$ to $\m{O}_{2^{2^k}}$. Finally, we identify a simplicial map $f:\m{P}_{[R]}^{(t-1)} \to \Phi(\m{P}_{[R]}^{(t)})$ that allows us to conclude that
\[
\delta' : \m{P}_{[R]}^{(t-1)} \to \m{O}_{2^{2^k}} 
\]
defined by 
\[
\delta' = \Phi(\delta)\circ f
\]
satisfies the hypotheses of Theorem~\ref{theo:main}, guaranteeing the existence of a $(t-1)$-round algorithm for $2^{2^k}$-coloring the ring. 

\begin{figure}[htb]
\begin{center}
\begin{tikzpicture}
\matrix (m) 
[matrix of math nodes,row sep=4em,column sep=4em,minimum width=2em] 
{
\m{P}^{(t)}_{[R]} &  \Phi(\m{P}^{(t)}_{[R]}) &  \m{P}^{(t-1)}_{[R]} \\
 \m{O}_k          &   \Phi(\m{O}_k)\subseteq \m{O}_{2^{2^k}}\\ 
}; 
\path[-stealth] 
(m-1-1) edge node [above] {$\Phi$} (m-1-2)
(m-2-1) edge node [above] {$\Phi$} (m-2-2)
(m-1-1)  edge node [left] {$\delta$} (m-2-1)
(m-1-3)  edge node [above] {$f$} (m-1-2)
(m-1-3)  edge node [above] {$\delta'$} (m-2-2)
(m-1-2) edge node [left] {$\Phi(\delta)$} (m-2-2);
\end{tikzpicture}
\end{center}
\vspace*{-5ex}
\caption{\sl Commutative diagrams in the proof of Corollary~\ref{cor:reproving-linial}. }
\label{fig:f**f**}
\end{figure}

\subparagraph{Detailed arguments.} 

Let us consider any complex $\m{K}$ with vertices $(p_i,v)$ with $i\in\{-1,0,1\}$, and $v\in V$ where $V$ is a finite set of values. Note that both $\m{O}_k$ and $\m{P}_{[R]}^{(t)}$ are of this form, where the values are respecitively colors in $\m{O}_k$, and views at distance~$t$ in~$\m{P}_{[R]}^{(t)}$. We define the functor~$\Phi$ as follows. The complex $\Phi(\m{K})$ is on the set of vertices $(p_i,\mathbf{S})$ where $\mathbf{S}=\{S_1,\dots,S_\ell\}$ for some $\ell\geq 0$, and $S_i\subseteq V$ for every $i=1,\dots,\ell$.  A set $\{(p_{-1},\mathbf{S}_{-1}),(p_{0},\mathbf{S}_{0}), (p_{1},\mathbf{S}_{1})\}$ forms a facet of $\Phi(\m{K})$ if for every $i\in\{0,1\}$, 
\begin{equation}\label{eq:f**f**}
\exists S \in \mathbf{S}_{i-1}\;  \; \forall S' \in \mathbf{S}_{i} \;\; \exists v'\in S' \; \; \forall v \in S \; : \; \{(p_{i-1},v),(p_i,v')\}\in\m{K}.
\end{equation}
Given a simplicial map $\psi:\m{A}\to\m{B}$ the map $\Phi(\psi)$ is defined as 
\[
\Phi(\psi)(p_i,\mathbf{S})=\Big ( p_i, \Big \{ \big \{ \pi_2 \circ \psi(p_i,v_{1,1}), \dots, \pi_2 \circ \psi(v_{1,s_1}) \big\}, \dots, \big\{ \pi_2 \circ \psi(v_{\ell,1}),\dots,\pi_2 \circ \psi(v_{\ell,s_\ell}) \big\} \Big\}\Big )
\]
for every $i=\{-1,0,1\}$, and every $\mathbf{S} = \{S_1,\dots,S_\ell\}$ with $S_j=\{v_{j,1},\dots,v_{j,s_j}\}$ and $s_j\geq 0$, where $\pi_2: \m{B} \to V$ is the mere projection $\pi_2(p_i,v)=v$ for every value~$v$. By construction, $\Phi(\psi):\Phi(\m{A})\to\Phi(\m{B})$ is simplicial. Note that if $\psi$ is name-preserving and name-independent, then so is $\Phi(\psi)$.

Next, we observe that $\Phi(\m{O}_k)$ is a sub-complex of $\m{O}_{2^{2^k}}$. To see why, note first that $\Phi$ maps vertices of $\m{O}_k$ to vertices of $\m{O}_{2^{2^k}}$. Moreover, a facet $F=\{(p_{-1},\mathbf{S}_{-1}),(p_{0},\mathbf{S}_{0}), (p_{1},\mathbf{S}_{1})\}$ of $\Phi(\m{O}_k)$ is a facet of $\m{O}_{2^{2^k}}$. Indeed, Eq.~\eqref{eq:f**f**} guarantees the existence of a set $S$ in $\mathbf{S}_{-1}$ such that for every set $S'$ in $\mathbf{S}_{0}$, there exists a color $v'$ in $S'$ that is different from all the colors in~$S$. It follows that $S\notin \mathbf{S}_{0}$, and therefore $\mathbf{S}_{-1}\neq \mathbf{S}_{0}$. By the same argument, $\mathbf{S}_{0}\neq \mathbf{S}_{1}$, and thus $F$ is a facet of $\m{O}_{2^{2^k}}$, as claimed. 

Finally, we define the simplicial map $f:\m{P}_{[R]}^{(t-1)} \to \Phi(\m{P}_{[R]}^{(t)})$ as follows. Let us consider a vertex $(p_i,w)\in\m{P}_{[R]}^{(t-1)}$, with 
\[
w=(z_{-(t-1)},\dots,z_{-1}, z_0, z_1,\dots,z_{t-1})\in [R]^{2t-1} \; \; \mbox{with} \;  z_{-(t-1)} < \dots < z_{t-1}.
\]
For every $b \in [R]$ with $b > z_{t-1}$, let 
$
W_i^{b}=\{ awb  :a \in [R], a < z_{-(t-1)} \}, 
$
and let 
\[
\mathbf{W}_i =\{W_i^{b} : b \in [R], b > z_{t-1}\}.
\]
We set $f(p_i,w)=(p_i,\mathbf{W}_i)$. This mapping maps every vertex of $\m{P}_{[R]}^{(t-1)}$ to a vertex of~$\Phi(\m{P}_{[R]}^{(t)})$. Let us show that $f$ is simplicial. For this purpose, let us consider a facet 
\[
F=\{(p_{-1},x'xw),(p_0,xwy),(p_1,wyy')\}
\]
of  $\m{P}_{[R]}^{(t-1)}$. Here $w=(z_{-(t-2)},\dots,z_{-1}, z_0, z_1,\dots,z_{t-2})\in [R]^{2t-3}$ with $x'<x< z_{-(t-2)} < \dots < z_{t-2}<y<y'$. 
We now show that the two sets $W_{-1}^{y} \in \mathbf{W}_{-1}$ and $W_{0}^{y'} \in \mathbf{W}_{0}$  witness the validity of Eq.~\eqref{eq:f**f**}, from which we conclude that $f(F)$ is a facet of $\Phi(\m{P}_{[R]}^{(t)})$. 
Consider $W_{-1}^{y} \in \mathbf{W}_{-1}$, let $W_{0}^{b} \in \mathbf{W}_{0}$, and let $x'xwyb \in W_{0}^{b}$. The view $ax'xwy$ for $p_{-1}$ is compatible with the view $x'xwyb$ for $p_0$, for every $a<x'$. Therefore, 
for every set $W_0^b \in \mathbf{W}_{0}$, there exists a view $x'xwyb \in W_0^b$ such that, for every view $ax'xwy \in W_{-1}^{y}$,
 \[
  \{(p_{-1},ax'xwy),(p_0,x'xwyb)\}\in \m{P}_{[R]}^{(t)}.
\]
Hence Eq.~\eqref{eq:f**f**} is satisfied for $p_{-1}$ and $p_0$. By the same arguments, using $W_{0}^{y'}$ instead of $W_{-1}^{y}$, Eq.~\eqref{eq:f**f**} is satisfied for $p_{-1}$ and $p_0$, from which it follows that $f(F)$ is a facet of $\Phi(\m{P}_{[R]}^{(t)})$. We conclude that $f$ is simplicial. 

Since both $f$ and $\Phi(\delta)$ are simplicial, the map $\delta'=\Phi(\delta)\circ f$ is simplicial too, which completes the proof by application of Theorem~\ref{theo:main}. 
\qed\medskip

By iterating Corollary~\ref{cor:reproving-linial}, we obtain that if there exists a $t$-round algorithm for 3-coloring~$C_n$, then there is a zero-round algorithm for coloring $C_n$ with a color pallet of $^{2t+2}2$ colors, where $^h2$ denotes the tower of exponentiels of height~$h$, from which the lower bound of $\frac12\log^*n-1$ rounds for 3-coloring $C_n$ follows.

\section{Conclusion and Further Work}
This paper shows that the study of algorithms for solving LCL tasks in the \local\/ model can be achieved by considering simplicial complexes whose sizes are independent of the number of nodes, and independent of the number of possible IDs that could be assigned to these nodes. We provide an application of our framework by providing a topological perspective of the lower bound proof for 3-coloring the $n$-node ring. Two main directions for further work can be identified. 

A first direction is understanding topological properties of the carrier map~$\Xi_t:\m{I}_{d,X}\to \m{P}_{d,X}^{(t)}$ occurring in the \local\/ model. This map governs the topology of the $t$-round protocol complexes $\m{P}_{d,X}^{(t)}$. It is known from the preliminary study in~\cite{CastanedaFPRRT21} that this topology heavily depends on the structure of the (class of) graph(s) $\m{G}$ in which the algorithm is supposed to be executed. However, still very little is known about how the elementary topological properties of the protocol complexes evolves from one round to the next. 

Another direction of research is understanding what governs the existence of the simplicial map $\delta:\m{P}_{d,X}^{(t)}\to \m{O}$ in Theorem~\ref{theo:main} (see also Figure~\ref{fig:commdiag}). In the shared memory setting, it is known that the existence of such a map  for consensus or $k$-set agreement tasks under the wait-free model is governed by the level of connectivity of the protocol complexes (i.e., the ability to contract high dimensional spheres). Would it be possible to provide similar types of characterization in the \local\/ model, say for tasks such as coloring or MIS? 
%

\bibliography{biblio-local-complexes}

\end{document}